\documentclass[twocolumn,twoside,10pt]{article}
\usepackage{flushend}
\usepackage[dvips]{graphics}
\usepackage{macro}
\usepackage[utf8]{inputenc}
\usepackage{graphicx}
\usepackage[colorlinks,allcolors=cyan!70!black]{hyperref}
\usepackage{bm}
\usepackage{amssymb}
\usepackage{amsbsy}
\usepackage{amsmath}
\usepackage{mathtools}
\usepackage{xcolor}
\usepackage{stmaryrd}
\usepackage{mathrsfs}
\usepackage{tikz}
\usetikzlibrary{shapes}
\usepackage{cancel}
\usepackage[normalem]{ulem}
\usepackage{cuted}
\usepackage{flushend}
\usepackage{chemformula} 
\usepackage[T1]{fontenc} 
\usepackage[super,numbers]{natbib}

\yearpublication{2025} \vol{X} \issueno{Y}
\authorhead{Mariya et al}

\newcommand*{\etal}{\textit{et al.}}

\DeclareUnicodeCharacter{0301}{\'{}}  

\begin{document}
\title{Subdiffusion of sticky dendrimers in an associative polymer network}

\author{Silpa Mariya}
\affiliation{IITB-Monash Research Academy, Indian Institute of Technology Bombay, Mumbai, 400076, India. \\Department of Chemical Engineering,
  Indian Institute of Technology Bombay, Mumbai, 400076, India. \\
  Department of Chemical and Biological Engineering, Monash University, Melbourne, VIC, 3800, Australia}
\author{Jeremy J. Barr}
\affiliation{School of Biological Sciences, Monash University, Clayton, VIC, 3800, Australia.}
\author{P. Sunthar}
\affiliation{Department of Chemical Engineering, Indian Institute of Technology Bombay, Mumbai, 400076, India.}
\author{J. Ravi Prakash$^*$}
\affiliation{Department of Chemical and Biological Engineering, Monash University, Melbourne, VIC, 3800, Australia \\
Email: ravi.jagadeeshan@monash.edu}
\date{\today}

\begin{abstract}
We investigate the static and dynamic properties of dendrimers diffusing through a network of linear associative polymers using coarse-grained Brownian dynamics simulations. Both dendrimers and network chains are modelled as bead-spring chain polymers, with hydrodynamic interactions incorporated for the accurate prediction of dynamic properties. Linear chains form a network via the associating groups distributed along their backbones, and the dendrimers interact attractively or repulsively with the network, enabling a direct comparison of sticky and non-sticky behaviour of dendrimers. Structural analysis reveals that while non-sticky dendrimers shrink with increasing network concentration, similar to linear polymer behaviour, sticky dendrimers exhibit stretching at low concentrations due to binding interactions. Dendrimer dynamics are largely insensitive to network architecture but are strongly influenced by the strength of dendrimer–network interactions. Increasing attraction to the network leads to subdiffusive motion and non-Gaussian displacement statistics, even when dendrimers are smaller than the average mesh size. The long-time diffusivity aligns with theoretical predictions for nanoparticle transport in polymer networks. Additionally, dendrimers deform the network locally, altering the mesh size distribution depending on their stickiness. These findings offer insight into the interplay between macromolecular architecture, binding interactions, and transport in polymeric environments.
\end{abstract}

\maketitle

\section{\label{sec:intro} Introduction}

\begin{figure*}[t]
	\begin{center}
		\begin{tabular}{cc}
			\resizebox{7cm}{!} {\includegraphics[width=9cm, height=7cm]{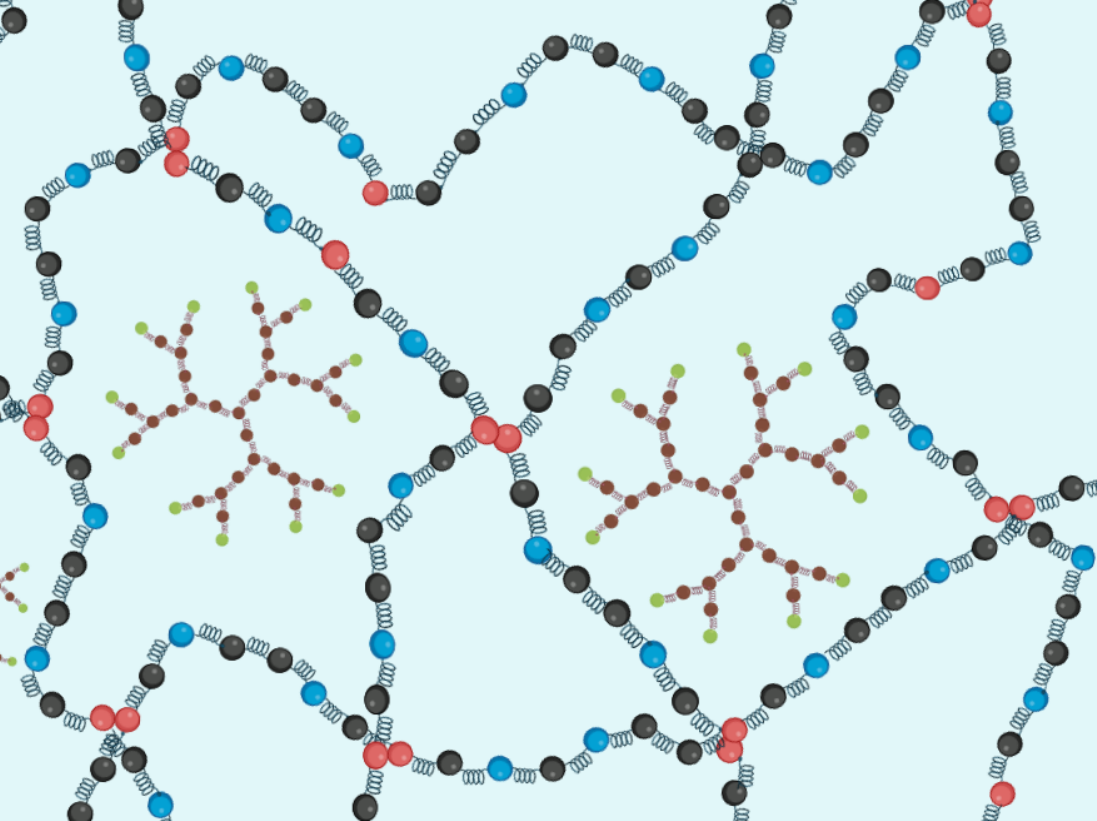}} &		
			\resizebox{5.9cm}{!} {\includegraphics[width=3.0cm]{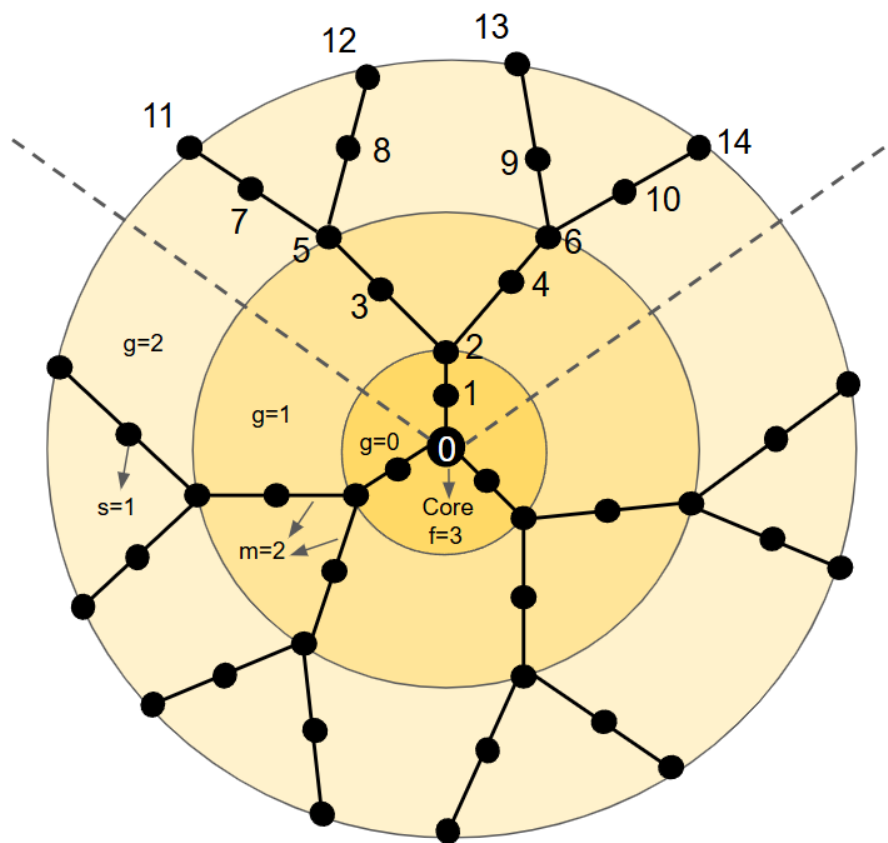}}\\
			(a) & (b)  
		\end{tabular}
	\end{center}
	\vspace{-10pt}
        \caption{\footnotesize (Color online) (a) Schematic representation of dendrimers in a network of linear chains. The black and brown beads are the homopolymer beads belonging to the linear chain and dendrimer, respectively. The red, blue and green beads are stickers. (b) A generation two ($g=2$) dendrimer with functionality three ($f=3$) and one spacer bead ($s=1$). The order of dendra ($m=f-1$) is two ($m=2$). Beads corresponding to each generation are arranged in concentric circles. The bead numbering scheme is explained in Section S1 in the Supporting Information.}
    \label{fig:dendrimer}
\end{figure*}

Anomalous diffusion of tracer particles in complex media is characterised by deviations from Brownian motion, where the mean squared displacement (MSD) has a power law dependence on time, $\textrm{MSD} = K t^{\alpha}$ where $K$ is the diffusion constant, $\alpha$ is the diffusion exponent with $\alpha \neq 1$~\cite{metzler2000random}. This behaviour can be subdiffusive ($\alpha<1$) or superdiffusive ($\alpha>1$), depending on the properties of the medium and the interaction between the medium and tracer. Subdiffusion has been widely observed in systems like particle transport in the cytoplasm,~\cite{norregaard2017manipulation} diffusion of proteins through the Nuclear Pore Complex (NPC),~\cite{chatterjee2011subdiffusion,ando2017cooperative} virus diffusion in mucus,~\cite{barr2013bacteriophage,barr2015subdiffusive} and nanoparticle diffusion in polymer solutions~\cite{nahali2018nanoprobe} and gels.~\cite{godec2014collective} While much of the existing research focuses on hard-sphere tracers, the diffusion of soft colloids remains less understood despite their increasing use in drug delivery and catalysis due to their tunable properties. In this research, we investigate the dynamics of dendrimers that are capable of interacting with an associative polymer network of linear chains (as shown in Figure~\ref{fig:dendrimer}(a)), to explore tracer diffusion through a polymer network.  

Polymers with highly branched symmetric architectures are called dendrimers. They consist of a central core, with dendrimeric arms or branches radiating from it. They are synthesised in a stepwise manner by adding layers of short chains to the ends of each arm, which will then constitute a new generation. The size, shape and properties of dendrimers are dependent on four parameters: functionality ($f$), generation number ($g$), spacer length ($s$), and the order of dendra ($m$). Figure~\ref{fig:dendrimer}(b) shows a second generation ($g=2$) symmetric dendrimer with functionality three ($f=3$), one spacer bead ($s=1$) and order of dendra two ($m=2$). Often referred to as soft colloids, dendrimers with low generation numbers have been shown to possess properties of linear chains,~\cite{bosko2011universal,mariya2024universal} while those of high-generation dendrimers are similar to hard spheres.~\cite{cheng2002diffusion} The transition between the two is governed by $f$, $s$, $g$ and $m$.~\cite{vlassopoulos2004colloidal,likos1998star,mariya2024universal}

Two major analytical theories have been developed to describe nanoparticle dynamics within permanently crosslinked and associative polymer networks. While the theories of hopping and sticky Rouse motion proposed by Cai \etal~\cite{cai2012structure} describes the diffusion of non-sticky hard spheres with a size larger than the network mesh size, the theory by Dell and Schweizer~\cite{dell2014theory} includes soft nanoparticles with attractive interactions with the network as well. Unlike in polymeric liquids, dry networks or gels do not exhibit constraint release, making nanoparticle mobility completely dependent on the hopping process. Hopping occurs when one of the network strands confining the nanoparticle slips around it due to thermal fluctuations.~\cite{cai2015hopping} The key parameter governing nanoparticle mobility is the confinement parameter, $C$. It is defined as the ratio of the nanoparticle diameter ($\sigma_{\mathrm{NP}}$) to the mesh size of the network (which is equal to the correlation length $\xi$ when entanglements are neglected), $C=\sigma_{\mathrm{NP}}/\xi$. The significance of the confinement parameter in determining the dynamics of nanoparticles has been independently predicted by Cai \etal~\cite{cai2015hopping} and Dell and Schweizer,~\cite{dell2014theory}, although the two theories offer qualitatively different predictions regarding nanoparticle diffusivity. Apart from diffusivity, several other properties have been studied for nanoparticles, including the diffusion exponent, the probability distribution function of displacements and velocity autocorrelation functions. Even though these theories have been developed for hard spheres, they have not been tested for soft colloids in polymer networks with varying interactions with the network. 

Due to the significance of tracer transport in various fields, numerous simulation studies have been conducted to understand the effect of matrix parameters on solute diffusion. Early models represented the network as an array of fixed obstacles~\citep{netz1997computer} while recent ones have incorporated aspects such as network connectivity, flexibility,~\citep{kumar2019transport} disorder and polydispersity.~\citep{sorichetti2021dynamics} Most of these models construct networks by placing crosslinks on a regular lattice~\citep{lu2021potential} and connecting them via simple springs,~\citep{zhou2009brownian} thereby neglecting the effect of strands connecting crosslinks, like entanglement and strand dynamics. Moreover, real hydrogels are polydisperse systems with disorder. Recent simulation studies have accounted for some of these factors and compared them with theoretical predictions.~\cite{kamerlin2016tracer,cho2020tracer,sorichetti2021dynamics,zhao2024coarse} However, they neglect hydrodynamic interactions, which have been shown to impact polymeric systems in the semidilute regime.  Additionally, there has been limited research to understand the transport of soft colloids, such as dendrimers, in polymer networks, with most studies focusing on hard spheres.~\cite{ge2024theoretical} 

To bridge this gap, we simulate dendrimers in an associative linear polymer network using Brownian dynamics simulations, incorporating key interactions such as excluded volume and hydrodynamic interactions. Building upon insights from our previous study,~\cite{mariya2024universal} we examine the static and dynamic properties of two dendrimer architectures as they diffuse within the network. By varying the linear polymer concentration and the interaction strengths, we analyse the size and bead density distribution of dendrimers. Furthermore, we examine the impact of interaction strength, sticker spacing, and linear polymer concentration on dendrimer diffusion. We also explore how dendrimers influence the mesh size of the polymer network. It is important to note that our model does not account for entanglements and is therefore not suitable for studying concentrated entangled systems, where hydrodynamic interactions are screened.%

The paper is organised as follows: The model, the governing equations and the various intra- and intermolecular interactions included in the model are discussed in section~\nameref{sec:Model}. The static and dynamic properties of sticky and non-sticky dendrimers are discussed in section~\nameref{sec:results}. The concluding remarks are given in section~\nameref{sec:conclusions}.

\section{\label{sec:Model} Materials and Method}

\subsection{ Model}

\begin{figure}[t]   
    \begin{center}
    {\includegraphics[width=7cm,height=!]{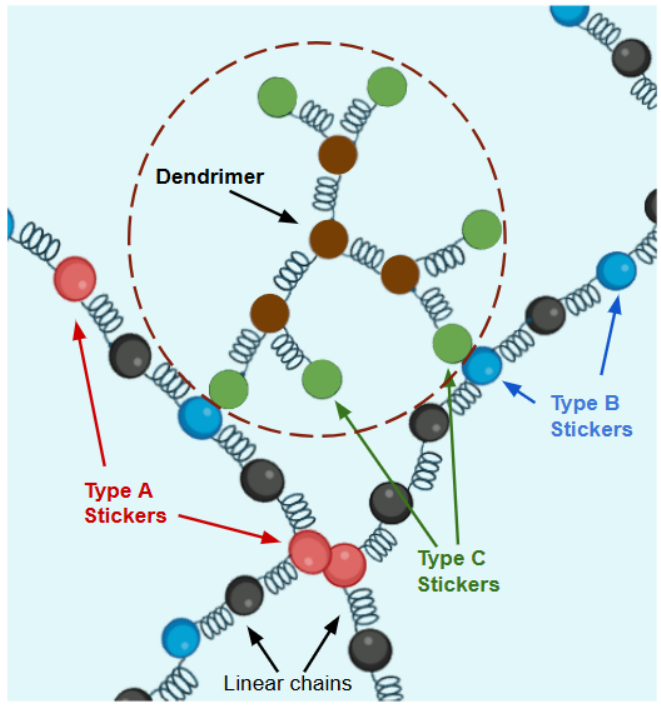}}
	\end{center}
	\vspace{-10pt}
	\caption[Types of stickers in the model]{\footnotesize Different types of stickers present in the model are shown. The red, blue and green beads are stickers of types A, B and C respectively. The black beads are the homopolymer beads present on the linear chain and the brown beads are those on the dendrimer, respectively.}
        \label{fig:sticker rules}
\end{figure}

The model employed in this research uses associative polymers to simulate both the tracer particle (i.e. the dendrimer) and the background polymer network. Associative polymers (APs), also called sticky polymers, are macromolecules capable of forming reversible physical bonds due to the presence of attractive/associative groups (or stickers).~\cite{rubinstein1997solutions,rubinstein2003polymer,winnik1997associative,santra2021universal,robe2024evanescent} The properties of associative polymers can be tuned by varying the number, strength and location of these attractive groups on the polymer. In this work, the background is a network of linear polymers that form inter and intra-chain associations through which the dendrimers diffuse. The arms of the dendrimers have stickers which interact with the linear chains. To efficiently simulate the network formation and adherence between dendrimers and linear chains, we include three types of stickers in the system as shown in Figure~\ref{fig:sticker rules}: 
\begin{enumerate}
    \item Type A: These are the network-forming stickers and are present only on the linear chains. They are allowed to stick to each other and are similar to crosslinkers in experiments. Only pairs of stickers are allowed to bind, and the binding of more than two stickers is prohibited.
    \item Type B: This type of sticker is interspersed between type A stickers on the linear chain. They can attach to the stickers on the dendrimers. In this case as well, only pairs of stickers are allowed to bind.
    \item Type C: They are present at the ends of the arms of dendrimers and stick only to type B stickers on the linear chains, one pair at a time.
\end{enumerate}
Thus, we only allow A-A and B-C interactions in the system, while all the other interactions are prevented. A multi-species Brownian dynamics simulation algorithm has been developed for this capable of simulating polymers with arbitrary architectures and sticky interactions.

\subsection{\label{sec:SM} Basic equations}

The dendrimers and linear chains in our model are represented as $N_{\textrm{b}}$ beads connected by  $N_{\textrm{b}}-1$ finitely extensible nonlinear elastic (FENE) springs.~\cite{Bird} The system contains $N_{\textrm{c}}^{\textrm{d}}$ dendrimer molecules with $N_{\textrm{b}}^{\textrm{d}}$ beads in a background of $N_{\textrm{c}}^{\textrm{lc}}$ linear chain molecules with $N_{\textrm{b}}^{\textrm{lc}}$ beads immersed in a Newtonian solvent, contained in a cubic simulation box of length $L$ and volume $V$, where $V=L^3$. In BD simulations, the chain configurations are given by the bead positions $\mathbf{r}_\mu\,(\mu = 1, 2, ..., N)$ which are obtained as a function of time by integrating the governing stochastic differential equation. The Euler integration algorithm form of this non-dimensionalized It$\hat{\textrm{o}}$ stochastic differential equation is given below: 
\begin{align}\label{gov-eqn}
\mathbf{r}_\mu(t + \Delta t) =\, & \mathbf{r}_\mu(t) + \frac{\Delta t}{4} \sum\limits_{\nu=1}^N\mathbf D_{\mu\nu}\cdot(\mathbf F_\nu^{\textrm{s}}+ \mathbf F_\nu^{\textrm{SDK}}) \notag \\
& +\frac{1}{\sqrt{2}}\sum\limits_{\nu=1}^N \mathbf B_{\mu\nu}\cdot\Delta \mathbf W_\nu
\end{align}
All length and time scales are non-dimensionalized using $l_H=\sqrt{k_BT/H}$ and $\lambda_H=\zeta/4H$, respectively, where  $k_B$ is the Boltzmann constant, $T$ is temperature, $H$ is the spring constant, and $\zeta=6\pi\eta_s a$ is the Stokes friction coefficient of a spherical bead with radius $a$ and $\eta_s$ is the solvent viscosity.

The bonded interaction between adjacent beads is modelled as a spring force, $\mathbf F_\nu^{\textrm{s}}$, using the FENE potential given by, 
\begin{align}\label{eq-fraenkel}
U_{\textrm{FENE}}= -\frac{1}{2}Q_0^2\ln \left( 1-\frac{r^{2}}{Q_0^2}\right)
\end{align}
Here $Q_0$ is the dimensionless maximum stretchable length of a single spring defined by $Q_0=\sqrt{b}$ where $b$ is FENE $b$-parameter. $k_B T$ is used to non-dimensionalize energy. 

The diffusion tensor $\pmb D_{\mu\nu}$ in Eq~\eqref{gov-eqn} is defined as $\pmb D_{\mu\nu} = \delta_{\mu\nu} \pmb \delta + \pmb \Omega_{\mu\nu}$, where $\delta_{\mu\nu}$ is the Kronecker delta, $\pmb \delta$ is the unit tensor, and $\pmb{\Omega}_{\mu\nu}$ is the hydrodynamic interaction tensor for which we use the regularized Rotne-Prager-Yamakawa (RPY) tensor given by:
\begin{equation}
{\pmb{\Omega}_{\mu \nu}} = {\pmb{\Omega}} ( {\mathbf{r}_{\mu \nu}})
\end{equation}
where ${\mathbf{r}_{\mu \nu}} = {\mathbf{r}_{\mu}} - {\mathbf{r}_{\nu}}={\mathbf{r}}$ and the function $\pmb{\Omega}$ is
\begin{equation}
\pmb{\Omega}(\mathbf{r}) =  {\Omega_1{ \pmb \delta} +\Omega_2\frac{\mathbf{r r}}{{r}^2}}
\end{equation}
with
\small{
\begin{equation*}
\Omega_1 = \begin{cases} \dfrac{3\sqrt{\pi}}{4} \dfrac{h^{\ast}}{r}\left({1+\dfrac{2\pi}{3}\dfrac{{h^{\ast}}^2}{{r}^2}}\right) & \text{for} \quad r\ge2\sqrt{\pi}h^{\ast} \\
 1- \dfrac{9}{32} \dfrac{r}{h^{\ast}\sqrt{\pi}} & \text{for} \quad r\leq 2\sqrt{\pi}h^{\ast} 
\end{cases}
\end{equation*}
and 
\begin{equation*}
\Omega_2 = \begin{cases} \dfrac{3\sqrt{\pi}}{4} \dfrac{h^{\ast}}{r} \left({1-\dfrac{2\pi}{3}\dfrac{{h^{\ast}}^2}{{r}^2}}\right) & \text{for} \quad r\ge2\sqrt{\pi}h^{\ast} \\
 \dfrac{3}{32} \dfrac{r}{h^{\ast}\sqrt{\pi}} & \text{for} \quad r\leq 2\sqrt{\pi}h^{\ast} 
\end{cases}
\end{equation*}
}
\noindent In the above equations, $h^{\ast}$, represents the hydrodynamic interaction parameter. It denotes the dimensionless bead radius in the bead-spring model and is defined as $h^{\ast} = a/(\sqrt{\pi k_BT/H})$.

The dimensionless tensor $\pmb{B}_{\mu\nu }$ is calculated from the decomposition of the diffusion tensor given by $\mathcal{B} \cdot {\mathcal{B}}^\text{T} = \mathcal{D} \label{decomp}$, where $\mathcal{D}$ and $\mathcal{B}$ are block matrices of $N \times N$ blocks each having dimensions of $3 \times 3$. The $(\mu,\nu)$-th block of $\mathcal{D}$ contains the components of the diffusion tensor $\pmb{D}_{\mu\nu }$, and the corresponding block of $\mathcal{B}$ is $\pmb{B}_{ \mu\nu}$. $\Delta\pmb W_\nu$ is a non-dimensional Wiener increment sampled from a real-valued Gaussian distribution with mean zero and variance $\Delta t$.

The next important non-bonded interaction existing between bead pairs is the excluded volume interaction. It could be between pairs of backbone monomers and/or stickers in the system. The term $\mathbf F_\nu^{\textrm{SDK}}$ in Eqn~\eqref{gov-eqn} gives the force experienced by bead $\nu$ due to this short-range interaction with its neighbouring beads modelled using the Soddemann-D\"{u}nweg-Kremer (SDK) potential, $U_{\text{SDK}}$, ~\cite{SDK}
\begin{align}\label{eq:SDK}
U_{\textrm{SDK}}=\left\{
\begin{array}{l }
\!\!\! 4\left[ \left( \dfrac{\sigma}{r} \right)^{12} - \left( \dfrac{\sigma}{r} \right)^6 + \dfrac{1}{4} \right] - \epsilon \,;  \,\, r\leq 2^{\frac{1}{6}}\sigma \vspace{0.5cm} \\
\!\!\! \dfrac{1}{2} \epsilon \left[ \cos \,(\alpha \left(\dfrac{r}{\sigma}\right)^2+ \beta) - 1 \right] ; \, 2^{\frac{1}{6}}\sigma \leq r \leq r_c \vspace{0.5cm} \\
\!\!\! 0; \quad  r \geq r_c
\end{array}\right.
\end{align}
Here $\sigma$ is the non-dimensional distance and is set to 1. The repulsive part of the SDK potential is described using a truncated Lennard-Jones (LJ) potential while the attractive component is a cosine function. The attractive well depth of the potential, $\epsilon$, determines the strength of interaction between bead pairs. $r_c$ is the cut-off radius, equal to $1.82 \sigma$ and the minimum of the potential is at $r=2^{1/6}\sigma$. These conditions have been used to determine the values of constants $\alpha$ and $\beta$.~\cite{santra2019universality} 

\begin{figure}[!h]   
    \begin{center}
    {\includegraphics[width=7cm,height=!]{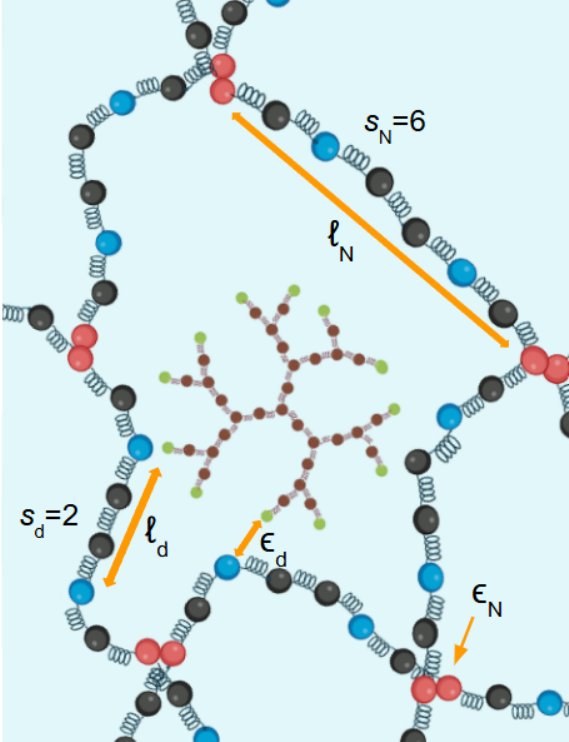}}
	\end{center}
	\vspace{-10pt}
	\caption[The various sticker-related parameters associated with the system of sticky dendrimer in a network]{\footnotesize The various sticker-related parameters associated with the system of sticky dendrimer in a network.}
        \label{fig:sticky dendrimer image}
\end{figure}

The SDK potential is employed to model both the sticker interactions as well as repulsion between non-sticky beads. This is done by varying the well depth of the potential. For purely repulsive interactions, we set $\epsilon=0$. The solvent quality decreases with increasing $\epsilon$. Unlike the conventional truncated Lennard-Jones (LJ) potential, the SDK potential has a short-ranged attractive tail that smoothly goes to zero at the cut-off distance, which increases the simulation efficiency~\cite{santra2019universality} making it a suitable choice for our simulations.

There are three important interaction parameters related to each sticker: 
\begin{itemize}
    \item Sticker strength: The strength of interaction between two stickers is determined by the well depth ($\epsilon$) of the SDK potential. A higher well depth implies a stronger sticker.
    \item Sticker functionality: The maximum number of stickers to which a reference sticker can adhere is referred to as its functionality. In this study, the functionality of all stickers is set to one.
    \item Sticker species: As mentioned in the model description, there are 3 types of stickers in our simulations (shown in Figure~\ref{fig:sticky dendrimer image}). The stickers that interact, i.e. A-A and B-C types, have a finite well depth for the SDK potential, $\epsilon_{\textrm{N}}$ and $\epsilon_{\textrm{d}}$ respectively, as shown in Figure~\ref{fig:sticky dendrimer image}. Here, the subscript 'N' represents network and 'd' represents dendrimers. The homopolymer beads and non-interacting stickers repel each other, hence the well depth of the potential, $\epsilon_{bb}=0$.
\end{itemize} 

The process of binding and unbinding of stickers is determined using a simple Monte Carlo (MC) process. This is applicable only in the case of sticker pairs A-A and B-C. When two unstuck stickers are within the cutoff radius, the energy change ($\Delta E$) due to a potential association is calculated. If a pseudo-random number, drawn between 0 and 1 from a uniform distribution, is less than the Boltzmann factor associated with energy change, $\exp(-\Delta E / k_B T)$, the stickers are allowed to bind. Bond breakage is also attempted similarly in the update sweep. If at least one of the 2 stickers within the cutoff radius is stuck already, new bond formation is not attempted. The MC and BD steps are carried out alternately and care has been taken to maintain detailed balance. This algorithm is similar to that developed by Robe \etal.~\cite{robe2024evanescent}

In this work, we have used a modified version of the GPU-accelerated Python package named HOOMD-Blue, developed at Michigan University ~\cite{anderson2020hoomd} for studying colloidal suspensions, along with the Positive Split Ewald algorithm (PSE)~\cite{fiore2017rapid} as a plugin for efficiently calculating the decomposition of the diffusion tensor. This has significantly brought down the computational cost of large many-body simulations with hydrodynamic interactions to $O(N \log N)$. Associative polymer interactions~\cite{robe2024evanescent} and branched polymer architecture~\cite{mariya2024universal} have been recently adapted to HOOMD-Blue based on an earlier in-house BD code using the Molecular Modelling ToolKit.~\cite{jain2012optimization,jain2015brownian,santra2021universal} 

\subsection{\label{sec:SD} Details of the simulation algorithm}

\begin{figure}[t]   
    \begin{center}
    {\includegraphics[width=7.5cm,height=6cm]{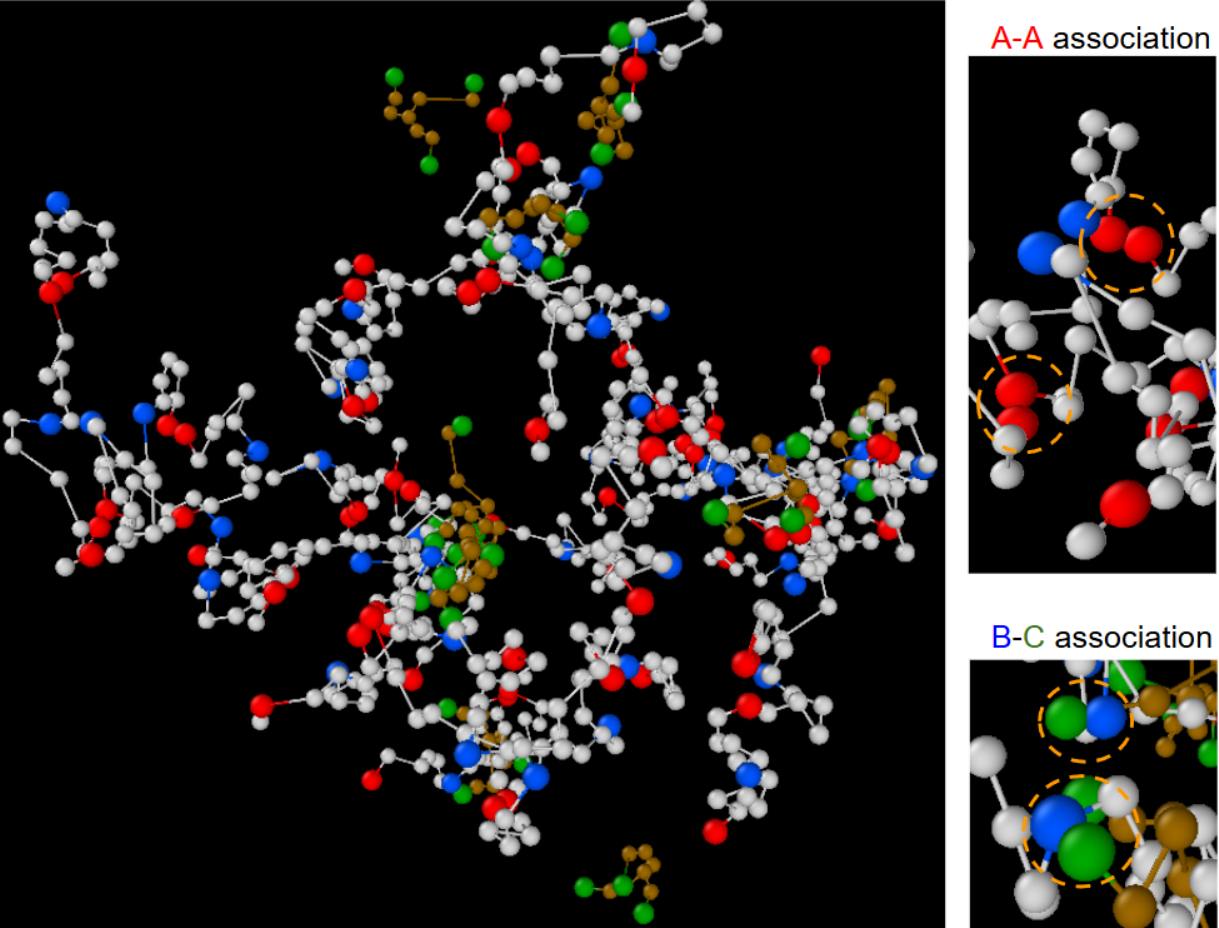}}
	\end{center}
	\vspace{-10pt}
	\caption[A snapshot from simulations]{\footnotesize A snapshot from simulations of sticky star polymers ($f,s,g,\chi$)=(3,2,0,0.5) in linear associative polymers at $c/c^{\ast}=0.5$. The white and brown beads in the system are the backbone monomers (non-sticky) on the linear chains and dendrimers respectively. The coloured beads are stickers of type A (red), type B (blue) and type C (green). Type A stickers interact with themselves to form a network while types B and C represent the linear chain-dendrimer interaction.}
        \label{fig:simulation snapshot}
\end{figure}

In this study, we have chosen two dendrimer architectures, a simple star polymer $(f,s,g)=(3,2,0)$ and a generation one dendrimer $(f,s,g)=(4,0,1)$. The order of dendra in both cases is given by $m=f-1$. The size ratio between the radius of gyration of dendrimers ($R_{\textrm{g0}}^{\textrm{d}}$) and linear chains ($R_{\textrm{g0}}^{\textrm{lc}}$) in the dilute limit ($\chi={R_{\textrm{g0}}^{\textrm{d}}}/{R_{\textrm{g0}}^{\textrm{lc}}})$) was chosen to be $0.5$. The length of linear chains was determined using the same methodology used in our previous work for dendrimers in a semidilute solution of linear polymers~\cite{mariya2024universal} (refer to Section~S1 in Supporting Information (SI) for more details). The length of the simulation box $L \geq 2R_{\mathrm{e}}$, where $R_{\mathrm{e}}$ is the end-to-end distance of linear chains in the solution so that molecules do not wrap around themselves. The concentration of the system refers to the monomer (or bead) concentration in the simulation box given by $c=N/V$, where $N$ is the total number of monomers given by $N = (N_{\textrm{c}}^{\textrm{lc}} \times N_{\textrm{b}}^{\textrm{lc}})+(N_{\textrm{c}}^{\textrm{d}} \times N_{\textrm{b}}^{\textrm{d}})$. The dendrimer concentration is always kept within the dilute limit and more linear chains are added to increase the overall concentration of the solution. The concentration is varied from dilute to $6c^{\ast}$, where $c^{\ast}$ is the overlap concentration defined in terms of the radius of gyration of the linear chains in the dilute limit ($R_{\textrm{g0}}^{\textrm{lc}}$), given by $c^{\ast} = N_{\textrm{b}}^{\textrm{lc}}/\left(({4}/{3})\pi \left(R_{\textrm{g0}}^{\textrm{lc}}\right)^3\right)$.

\begin{figure*}[t]
	\begin{center}
		\begin{tabular}{cc}
			\resizebox{8.0cm}{!} {\includegraphics[width=7.0cm]{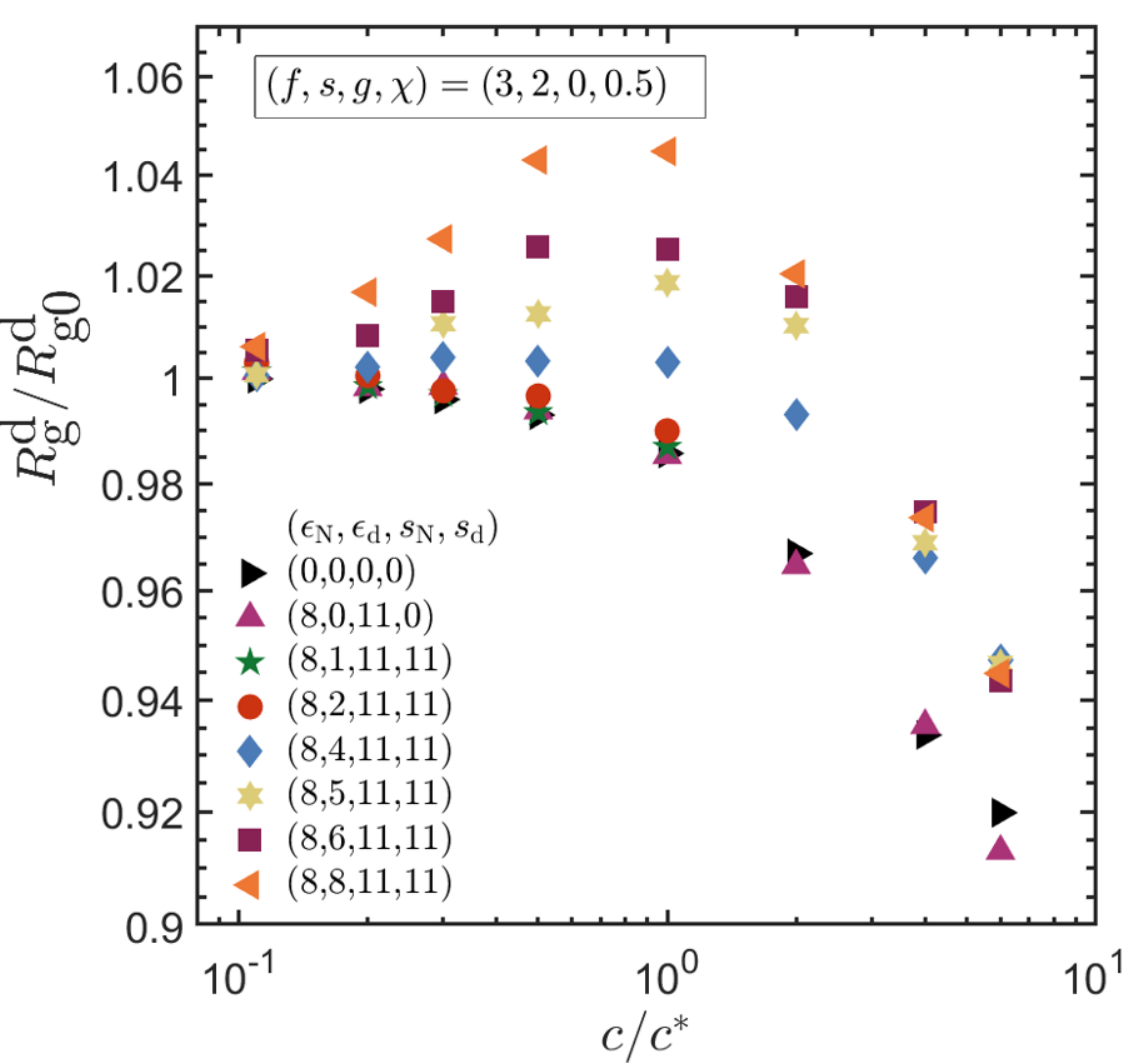}} &
                \resizebox{8.0cm}{!} {\includegraphics[width=7.0cm]{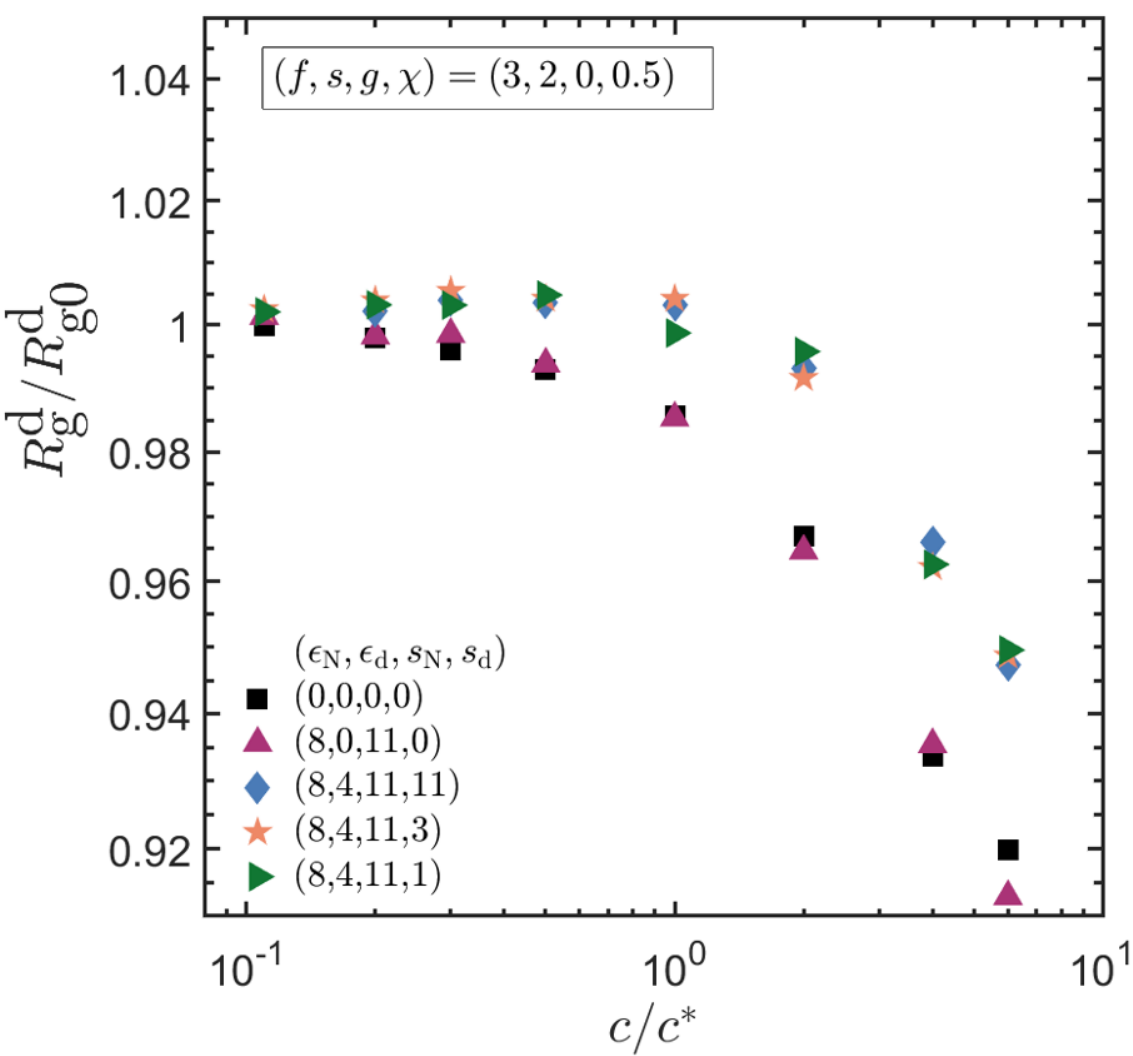}}\\
			(a) & (b)
		\end{tabular}
	\end{center}
		\vspace{-10pt}
        \caption[Radius of gyration of sticky dendrimers]{\footnotesize (a) Effect of the sticker strength and concentration on the normalised radius of gyration of dendrimers. The right pointing and up pointing triangles are non-sticky dendrimers in solution and network respectively. The stars and circles are sticky dendrimers with low sticker strengths, while the diamonds, hexagons, squares and left pointing triangles are for higher sticker strengths. (b) Effect of distance between type B stickers ($s_{\mathrm{d}}$) and concentration on sticky dendrimers with $\epsilon_{\textrm{N}}=8$ and $\epsilon_{\textrm{d}}=4$. For comparison, data for homopolymers in semidilute solutions is also included. Error bars are smaller than the marker size.}
    \label{fig:Rg of sticky dendrimers}
\end{figure*}

The inclusion of stickers also introduces two important length scales (shown in Figure~\ref{fig:sticky dendrimer image}): $l_{\textrm{N}}$ which is the distance between two A-type stickers and $l_{\textrm{d}}$ which is the distance between two B-type stickers along the backbone of the linear chain. This can also be represented in terms of the number of beads between 2 stickers, $s_{\mathrm{N}}$ and $s_{\mathrm{d}}$ for sticker types A and B, respectively. Decreasing the distance between stickers in turn increases the number of stickers per chain, $n_{\mathrm{N}}$ and $n_{\mathrm{d}}$. The details of the methodology to choose these quantities are given in Section~S2 of the SI. The sticker energies used in this study are $\epsilon_{\textrm{N}}= 0,1,4,6,8,12$ and $\epsilon_{\textrm{d}}=0,1,2,4,5,6,8$. The values of $s_{\mathrm{N}}$ and $s_{\mathrm{d}}$, and corresponding values of $l_{\textrm{N}}$ and $l_{\textrm{d}}$ are reported in Table~S2. The C-type stickers are present at the ends of the dendrimer arms. The interaction between any beads other than A-A and B-C stickers is purely repulsive and the well-depth of the SDK potential is zero. The dendrimer is referred to as `non-sticky' when $\epsilon_{\textrm{d}}=0$. 

The hydrodynamic interaction parameter is set at $h^{\ast}=0.2$ and the maximum stretchable length of the FENE spring $Q_0^2 = 50.0$. Each simulation trajectory has an equilibration phase ($\approx 10 \tau_{R}$) followed by a production run ($\approx 8 \tau_{R}$), where $\tau_{R}$ is the Rouse relaxation time given by $\tau_{R}= 1 / \left(2\sin^2\left(\pi/2N_{\textrm{b}}^{\textrm{lc}}\right)\right)$.~\cite{bird1987dynamics} The time step of integration is  $\Delta t= 10^{-3} \ \text{to} \ 10^{-4}$ in non-dimensional units. Properties were calculated from the ensemble averages of $700-3000$ independent trajectories. Figure~\ref{fig:simulation snapshot} shows a snapshot of a simulation box containing star polymers and linear chains at $c/c^{\ast} = 0.5$.

\begin{figure*}[tbph]
	\begin{center}
		\begin{tabular}{ccc}
			\resizebox{8.0cm}{!} {\includegraphics[width=7.0cm]{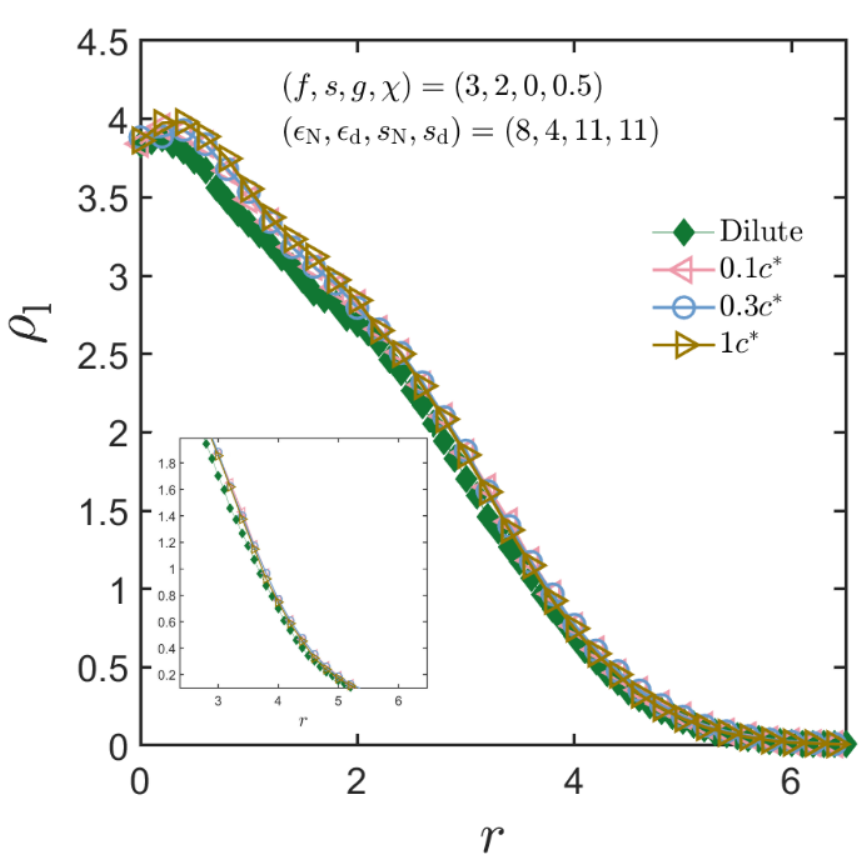}} &
                \resizebox{8.1cm}{!} {\includegraphics[width=7.0cm]{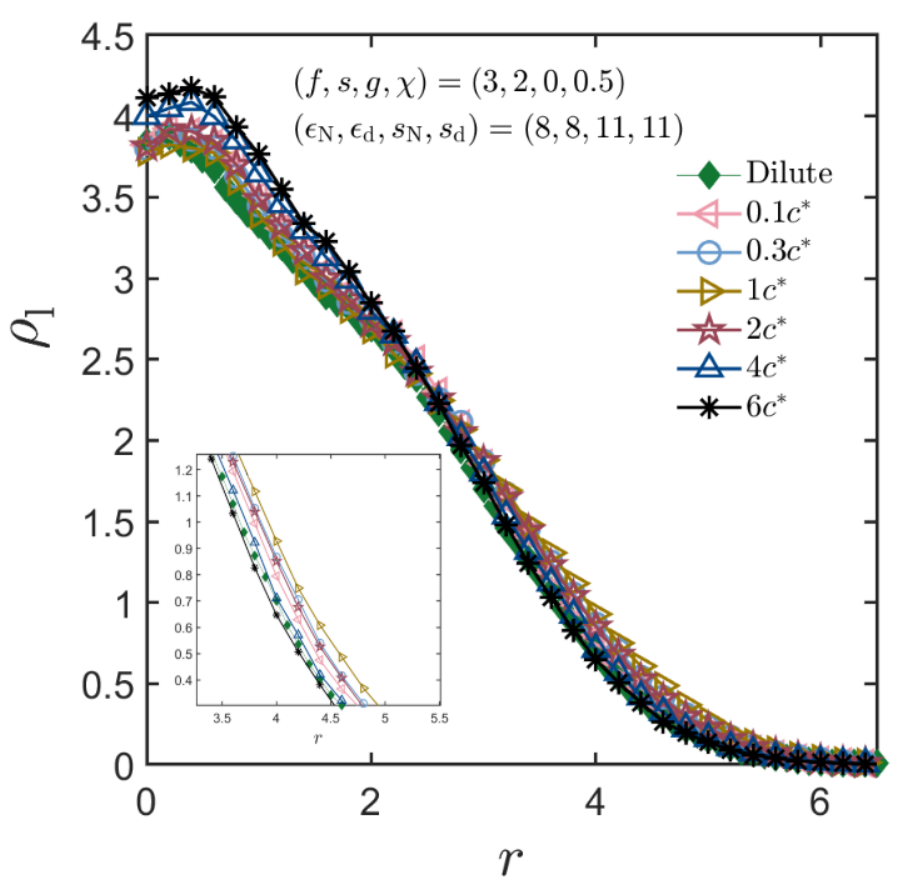}}\\
			(a) & (b)
		\end{tabular}
	\end{center}
\vspace{-10pt}
        \caption[Internal bead density of sticky dendrimers]{\footnotesize (a) Effect of the concentration on the internal bead density of sticky dendrimers for $\epsilon_{\textrm{d}}=4$ (a) and $\epsilon_{\textrm{d}}=8$ (b). The system is a functionality three star polymer with $(\epsilon_{\textrm{N}},s_{\mathrm{N}},s_{\mathrm{d}})=(8,11,11)$. Inset: Zoomed in version to show the difference in bead densities away from the core.}
    \label{fig:Internal density of sticky dendrimers}
\end{figure*}

\section{\label{sec:results}Results and discussion}

\subsection{Radius of gyration}

The confinement parameter $C$, emerges as the key factor governing the mobility of nanoparticles in polymer networks. Its significance has been independently established through theoretical predictions by by Cai \etal~\cite{cai2015hopping} and Dell and Schweizer.~\cite{dell2014theory} For a nanoparticle of fixed size, the confinement parameter can be modulated solely by varying $\xi$ of the network, where $\xi$ is the correlation length estimated using the scaling law~\cite{bird1987dynamics} $\xi=R_{\textrm{g0}}^{\textrm{lc}}\left(\dfrac{c}{c^{\ast}}\right)^{-0.75}$. However, dendrimers exhibit more complex behaviour, their size decreases along with $\xi$ with increasing concentration.  Our previous work~\citep{mariya2024universal} demonstrated that the subdiffusive behaviour of dendrimers in semidilute polymer solutions arises only when their radius of gyration exceeds the correlation length of the solution. Due to its influence on the dynamics, it is crucial to analyse the size of sticky and non-sticky dendrimers in polymer networks. 

\begin{align}
\label{Rg}
R_{\textrm{g}}^2 &=  \dfrac{1}{N_{\textrm{b}}} \langle \sum\limits_{\mu=1}^{N_{\textrm{b}}} \left( \mathbf{r}_\mu - \mathbf R_{\textrm{CM}} \right)^2 \rangle \ ; \ N_{\textrm{b}} \in \{ N_{\textrm{b}}^{\textrm{d}},N_{\textrm{b}}^{\textrm{lc}} \}
\end{align} 
where $R_{\textrm{CM}}$ is the centre of mass of the molecule given by
\begin{align}
    \mathbf R_{\textrm{CM}}=\dfrac{1}{N_{\textrm{b}}} \sum\limits_{\mu=1}^{N_{\textrm{b}}} \mathbf{r}_\mu 
\end{align}

Figure~\ref{fig:Rg of sticky dendrimers} shows the effect of sticker strength ($\epsilon_{\mathrm{d}}$) and the distance between type B stickers on linear chains ($s_{\mathrm{d}}$) on the size of dendrimers in a polymer network. The radius of gyration is normalised by that in the dilute case ($R_{\textrm{g0}}^{\textrm{d}}$). For comparison, this ratio for the same dendrimer architecture in a semidilute solution of linear chains without the network forming stickers is also included (black right-pointing triangles denoted as (0,0,0,0)). This was found previously to follow the scaling law for linear chains when expressed in terms of $c/c^{\ast}_{\mathrm{d}}$,~\citep{mariya2024universal} where $c^{\ast}_{\mathrm{d}}$ is the overlap concentration calculated with respect to the dendrimer. The size of a non-sticky dendrimer (magenta upward-pointing triangles denoted as (8,0,1,0)) and sticky dendrimers with low sticker strengths ($\epsilon_{\textrm{d}} = 1 \, \textrm{and} \, 2$) (star and circle symbols respectively) decreases with increasing concentration similar to the non-sticky dendrimer in a solution.

These results indicate that the conformational state of the background chains, whether they form a solution or a network, does not significantly influence the size of the non-sticky dendrimer. Instead, it is the local density of surrounding beads that plays a determining role. In the regime of weak sticker interactions (i.e., low values of $\epsilon_{\textrm{d}}$), the dendrimer undergoes frequent binding and unbinding events with the network, behaving similarly to a non-sticky dendrimer. As $\epsilon_{\textrm{d}}$ increases, however, the binding events become longer-lived,~\citep{robe2024evanescent} causing the dendrimer to remain attached to the network for extended periods. At low concentrations of linear chains, the availability of type B stickers is limited due to the sparsity of chains, and the larger average distance between stickers causes the dendrimer arms to stretch when they remain bound. This stretching leads to an increase in the normalised radius of gyration, reaching a maximum around $c/c^{\ast} = 1$, as shown in Figure~\ref{fig:Rg of sticky dendrimers}(a) (see the diamond, hexagon, square, and right-pointing triangle markers). Beyond this concentration, the linear chain crowding reduces the average distance between available attachment points. This, combined with the screening of excluded volume interactions, results in a decrease in the dendrimer size. Notably, the normalised radius of gyration of dendrimers with stronger sticker interactions (higher $\epsilon_{\textrm{d}}$) collapse onto a universal curve as concentration increases. 

In Figure~\ref{fig:Rg of sticky dendrimers}(b), the effect of decreasing $s_{\mathrm{d}}$ on the normalised size of a sticky dendrimer is plotted as a function of concentration. The values of $\epsilon_{\textrm{N}}$ and $\epsilon_{\textrm{d}}$ are fixed at $8$ and $4$ respectively. The decrease in distance between dendrimer stickers on linear chains gives more available points of attachment to the dendrimer. However, it seems not to affect the dendrimer size.

\subsection{Internal density distribution}

The internal structure of the molecule can be characterized by examining the distribution of beads relative to its center of mass. To quantify this, we calculate the linear bead density by binning beads along the major axis of the gyration tensor. Prior to this, all polymer configurations in the simulation ensemble are aligned along their respective major axes to ensure consistency. The linear bead number density, $\rho_{\rm{l}}(r)$, is defined as:
\begin{align}
    \rho_{\rm{l}}(r)=\dfrac{n_{\textrm{b}}(x+ \Delta x)-n_{\textrm{b}}(x)}{\Delta x}
\end{align}
where $n_{\textrm{b}}(x)$ is the number of beads of a molecule within a distance $x$ along the major axis, and $\Delta x$ is the length of the fixed interval. 

\begin{figure*}[t]
	\begin{center}
		\begin{tabular}{cc}		    
			\resizebox{8.0cm}{!} {\includegraphics[width=4cm]{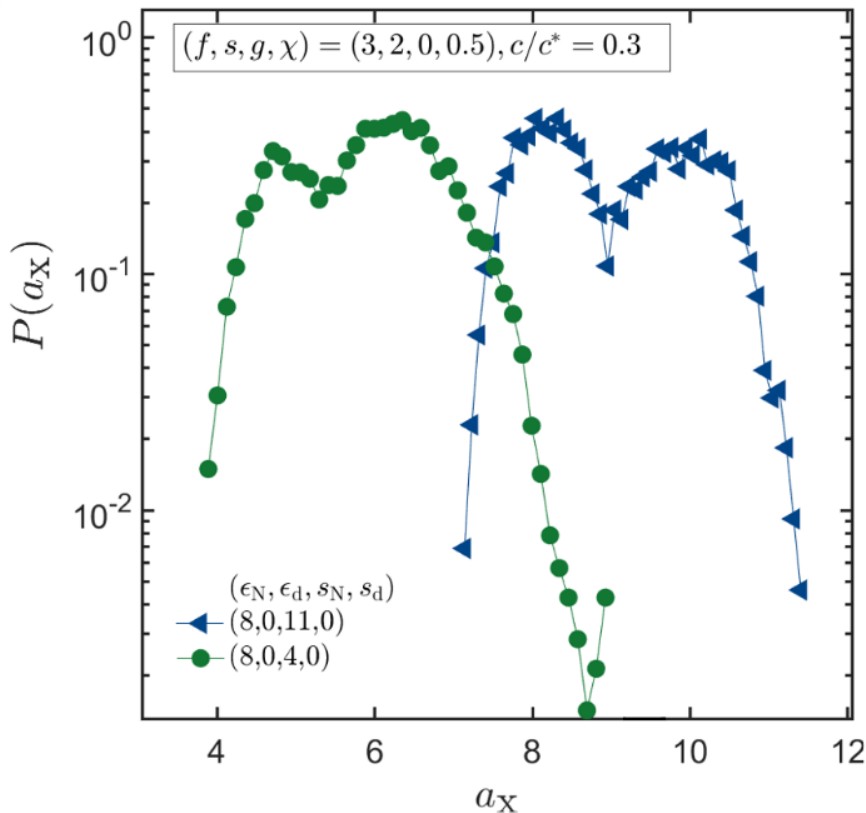}} &
			\resizebox{7.8cm}{!} {\includegraphics[width=4cm]{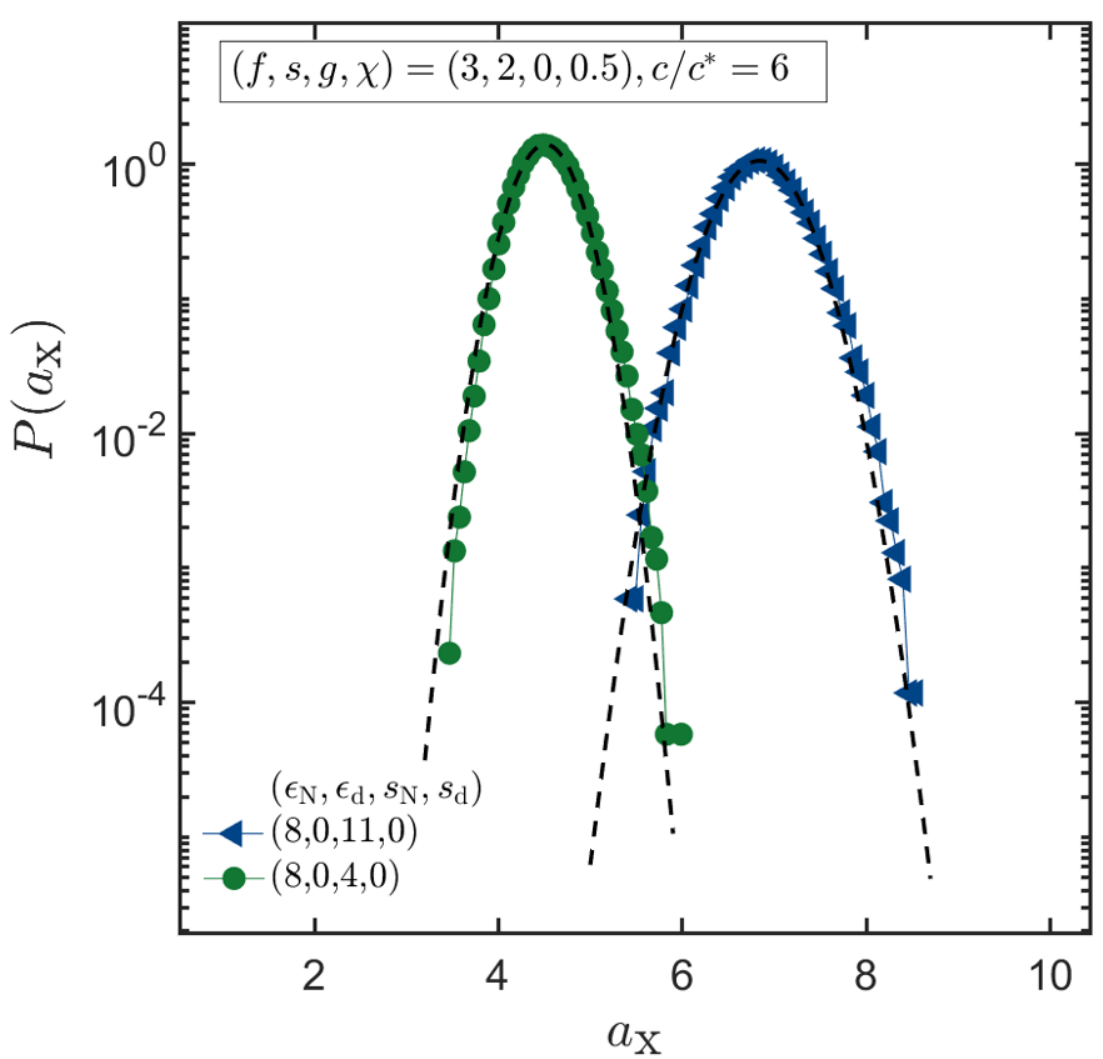}}\\
			(a) & (b)  \\
		\end{tabular}
	\end{center}
	\vspace{-10pt}
        \caption[Mesh size distribution with non-sticky dendrimers]{\footnotesize Effect of the distance between stickers and concentration on the mesh size distribution of the network. The system considered is for $\epsilon_{\textrm{N}}=8$ with a star polymer with functionality three in it. Results are shown for two values of $s_{\mathrm{N}}$ in each figure: $s_{\mathrm{N}}=11 \, \textrm{and} \, 4$. (a) and (b) correspond to $c/c^{\ast}=0.3 \, \textrm{and} \, 6$ respectively. The dashed curve in (b) are Gaussian fits to the data.}
    \label{fig:Mesh size distribution - non sticky}
\end{figure*}

\begin{figure*}[h!]
	\begin{center}
		\begin{tabular}{cc}
                \resizebox{8.0cm}{!} {\includegraphics[trim=140 10 140 10, clip,width=4.0cm]{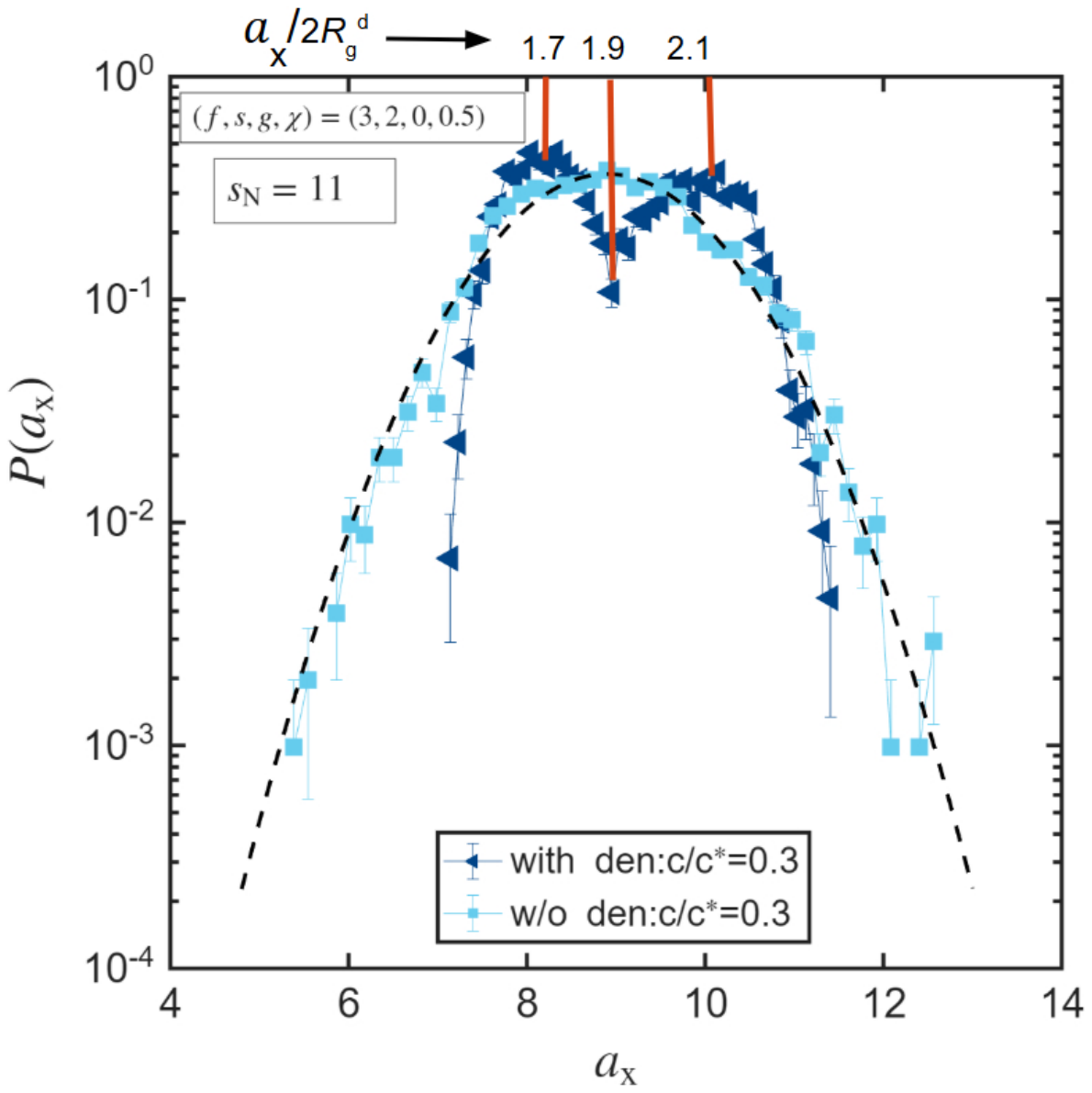}} &   
                \resizebox{5.0cm}{!} {\includegraphics[trim=50 20 80 20, clip,width=4.0cm]{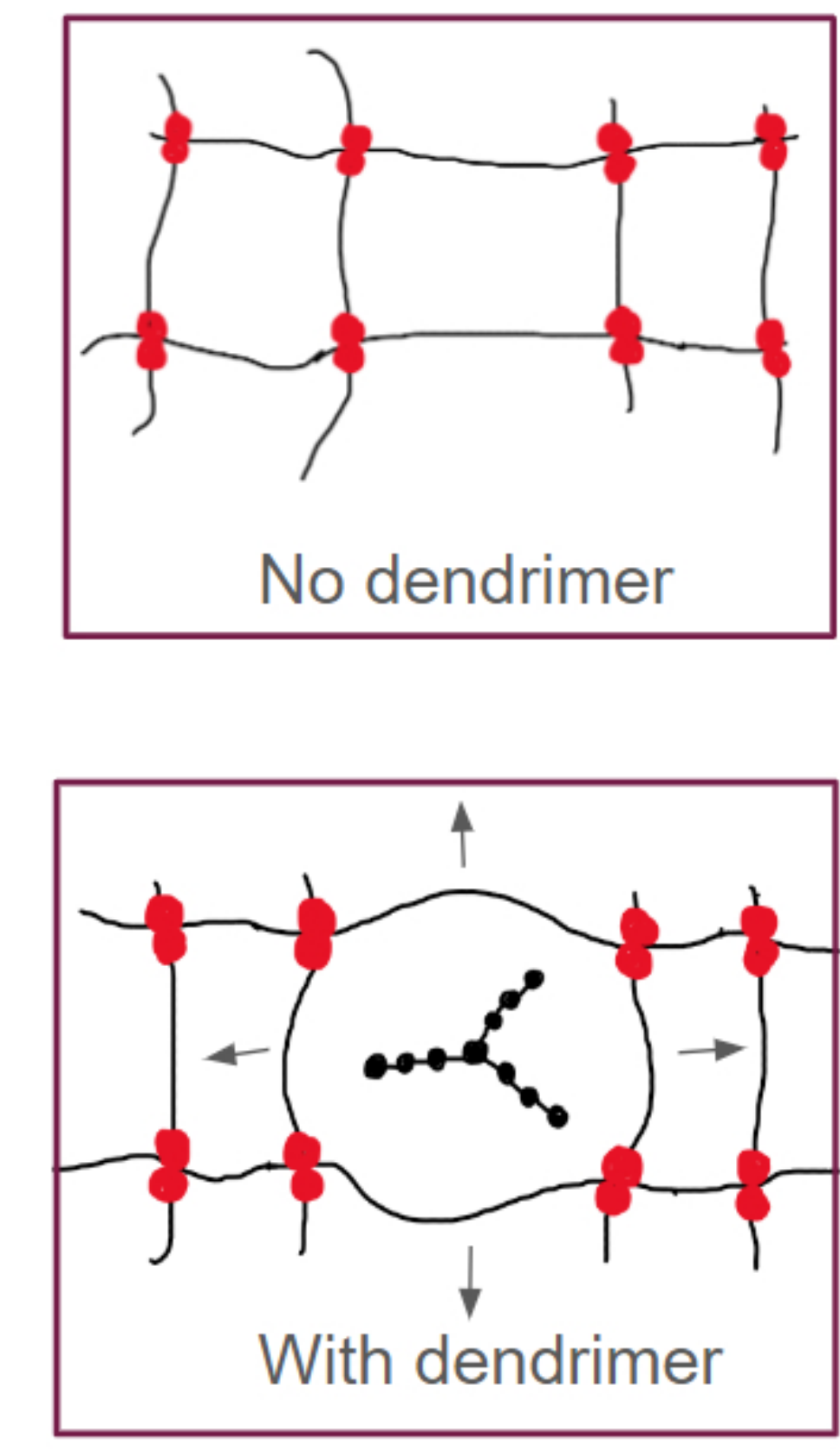}}\\			 
		\end{tabular}
	\end{center}
	\vspace{-10pt}
        \caption[Mesh size distributions with non-sticky dendrimers ]{\footnotesize Effect of the presence of non-sticky dendrimers on the mesh size distribution for $s_{\mathrm{N}}=11$ at $c/c^{\ast}=0.3$. The dashed curve is a Gaussian fit to the data and the vertical lines are guides to the eye for the peaks and minima in the distribution with respect to the size of the dendrimer. An illustration of the effect of dendrimers on the mesh size is shown on the right.}
    \label{fig:Mesh size distribution - with and without dendrimer}
\end{figure*}

The plot for the radius of gyration for dendrimers has shown that their size increases with increasing strength of adherence to the network due to a higher binding time. This observation suggests a potential decrease in internal density, particularly near the dendrimer core, as the arms become more extended. Figure~\ref{fig:Internal density of sticky dendrimers} shows the internal bead distribution along the major axis of dendrimers with two different sticker interaction strengths: $\epsilon_{\textrm{d}} = 4$ in panel (a) and $\epsilon_{\textrm{d}} = 8$ in panel (b), with fixed parameters $(\epsilon_{\textrm{N}}, s_{\mathrm{N}}, s_{\mathrm{d}}) = (8, 11, 11)$. In both cases, dendrimers exhibit a pronounced density peak at the core, indicating the presence of a dense central region regardless of interaction strength. For the weaker interaction case ($\epsilon_{\textrm{d}} = 4$), the internal bead density remains largely unaffected by increasing concentration up to $c/c^{\ast} = 1$, as shown in Figure~\ref{fig:Internal density of sticky dendrimers}(a). In contrast, at higher interaction strength ($\epsilon_{\textrm{d}} = 8$), where the radius of gyration exhibits a maximum near $c/c^{\ast} = 1$, a clear increase in bead density is observed in the outer regions of the dendrimer up to this concentration (Figure~\ref{fig:Internal density of sticky dendrimers}(b)). This redistribution is consistent with the swelling observed in $R_{\textrm{g}}^{\textrm{d}}$. Beyond $c/c^{\ast} = 1$, the dendrimer begins to contract, and the core density increases again.

\subsection{Distribution of mesh size } 

Both sticky and non-sticky dendrimers in our simulations encounter hindrance due to the cages formed by the polymer network, and the size and lifetime of these cages are influenced by the concentration of the network-forming linear chains, the sticker strength ($\epsilon_{\textrm{N}}$), and the spacing between stickers ($s_{\mathrm{N}}$) of type A. Even though $s_{\mathrm{N}}$ is the distance between stickers, it is not an estimate of the mesh size of the network. A more accurate measure can be obtained by calculating the mean spatial distance between bound stickers, which is denoted here by $a_{\textrm{x}}$. The quantity $a_{\textrm{x}}$ is defined as the mean squared distance between two adjacent stickers of type A that belong to the same linear chain and are stuck to a sticker on another chain. In other words, it is the distance between two stuck stickers on a linear chain that form an inter-chain association. The algorithm used to calculate $a_{\textrm{x}}$ is discussed in Section~S3 of the Supporting Information. The mesh size ($a_x$) of the network through which a probe particle diffuses affects the latter's dynamics to a great extent.~\cite{cai2015hopping,dell2014theory} However, it is equally crucial to understand the influence of these particles on the mesh size. Therefore, we calculated the distribution function of mesh sizes in a network with dendrimers. Figure~\ref{fig:Mesh size distribution - non sticky} presents the mesh size distribution in the network containing non-sticky dendrimers at two different concentrations ($c/c^{\ast}=0.3$ in ~\ref{fig:Mesh size distribution - non sticky}(a) and $c/c^{\ast}=6$ in ~\ref{fig:Mesh size distribution - non sticky}(b)) and each plot includes two values of $s_{\mathrm{N}}$. For additional concentrations and $s_{\mathrm{N}}$ values, refer to Section~S4 in the Supporting Information. At lower concentrations, the distribution of mesh size exhibits bimodality, with the local minima being more pronounced when $s_{\mathrm{N}}=11$ compared to the other values. As concentration increases, the distribution becomes unimodal and trends toward a Gaussian profile for all values of $s_{\mathrm{N}}$. 

To have a better understanding of the influence of dendrimers on the mesh size, we calculated the mesh size distribution for a network at similar concentrations, but without dendrimers. Figure~\ref{fig:Mesh size distribution - with and without dendrimer} shows the distribution for $s_{\mathrm{N}}=11$ at $c/c^{\ast}=0.3$. In the absence of dendrimers, the bimodality is absent and the distribution is Gaussian-like with a mean value similar to that of the network with dendrimers. The two peaks of the distribution are symmetrically spaced about the mean value of the no-dendrimer case, and the mean mesh size is much more than the diameter of the dendrimer calculated as $2R_{\textrm{g}}^{\textrm{d}}$. The vertical lines on the upper $x$-axis in Figure~\ref{fig:Mesh size distribution - with and without dendrimer} indicates that the maximas occur at $a_{\textrm{x}}/2R_{\textrm{g}}^{\textrm{d}} = 1.7 \, \textrm{and} \, 2.1$, while the mean mesh size is equal to $1.9$ times $2R_{\textrm{g}}^{\textrm{d}}$. The dashed curve is a Gaussian fit to the data.

It is important to note that at low concentrations, the number of dendrimers and linear chains is comparable. When a non-sticky dendrimer occupies a `cage' in the mesh, the strands forming the cage expand due to repulsion from the dendrimer molecule, thus increasing the mesh size as shown in the schematic in Figure~\ref{fig:Mesh size distribution - with and without dendrimer}. Consequently, the adjacent cages are likely to be compressed (as illustrated), leading to a higher probability of observing both larger and smaller meshes than the mean mesh size, thereby generating a bimodal distribution. However, as concentration increases, there are more stickers (Type A) available in the system, increasing their probability of binding to form a network with a large number of empty meshes, which are smaller. This is because the number of dendrimers, which remain dilute, has not changed. Therefore, the bimodality is absent at $c/c^\ast=6$ and the maxima of the distribution have shifted to the left (Figure~\ref{fig:Mesh size distribution - non sticky}(b)). This interpretation is further confirmed by simulations in which the number of dendrimers was increased, keeping the number of linear chains fixed (discussed in detail in Figure~S4 in Supporting Information). As shown there, the double peaks vanish as the number of dendrimers is increased, while the dynamics of dendrimers are unaffected by the presence of bimodality in the distribution.

\begin{figure}[tbph]
	\begin{center}
                \resizebox{8.0cm}{!} {\includegraphics[trim=120 10 120 0, clip,width=4.0cm]{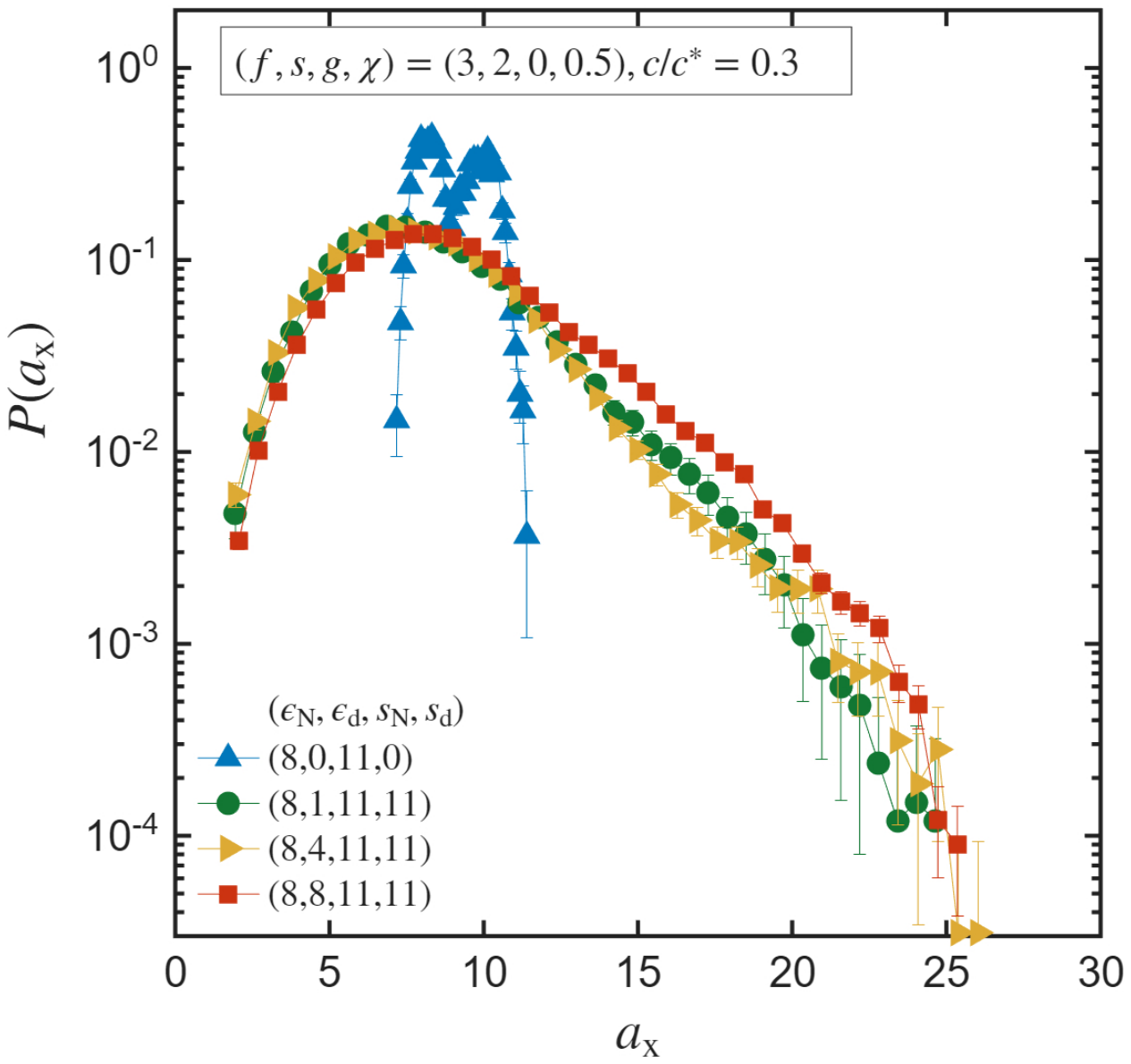}}
	\end{center}
	\vspace{-10pt}
        \caption[Mesh size distributions with a sticky dendrimer]{\footnotesize Effect of sticker strength $\epsilon_{\textrm{d}}$ on the mesh size distribution of the network. A range of $\epsilon_{\textrm{d}}$ is considered with a fixed value of $\epsilon_{\textrm{N}}=8$.}
    \label{fig:Mesh size distribution - sticky}
\end{figure}

The presence of a sticky dendrimer affects the network mesh size distribution; however, unlike the non-sticky case, it does not result in a bimodal distribution, as shown in Figure~\ref{fig:Mesh size distribution - sticky}. Instead, the distribution shifts and broadens, reflecting an increase in both the mean and the variability of mesh sizes. In the non-sticky case, the interaction between the dendrimer and the network is purely repulsive. In contrast, a sticky dendrimer interacts attractively, effectively pulling the surrounding chains toward itself and locally deforming the network. This leads to a wider range of mesh sizes, with increased probabilities of both smaller and larger pores, indicating greater heterogeneity. While the overall shape of the distribution remains qualitatively similar for different sticker strengths $\epsilon_{\textrm{d}}$, statistical tests reveal significant differences in the tails at large mesh sizes (discussed in Section~S5 in the Supporting Information). This suggests that stronger dendrimer-network attraction enhances the likelihood of rare, large pores through sustained local pulling.

Even though the methodology used is different, Sorichetti \etal~\cite{sorichetti2021dynamics} analysed the pore size distribution, $P(2r)$, of polydisperse polymer networks with spherical nanoparticles. Here, $r$ is the radius of the largest sphere without touching the network containing a randomly sampled point. The authors have considered repulsive and attractive nanoparticles referred to as RNPs and ANPs, respectively. In both cases, they observed double peaks in $P(2r)$, with the first peak being more pronounced and located at the same $r$ as that of a pure system with no nanoparticles, while the second peak is a smaller one. Specifically, RNPs were found to broaden the distribution, with the second peak located at $\sigma_{\mathrm{NP}} + \delta$, where $\sigma_{\mathrm{NP}}$ is the diameter of the nanoparticle and $\delta$ is a small shift. This is attributed to a larger effective diameter of the RNP due to its repulsion, which also leads to a deformation of the surrounding network. On the other hand, the attractive nanoparticles have a sharp peak at $\sigma_{\mathrm{NP}}$ due to the `cavity' it forms in the network. For smaller ANPs, they also observe a decrease in the mean mesh size due to the contraction of the network, a consequence of the attractive monomer-NP interaction. Their observations for systems with RNPs are similar to the non-sticky dendrimers cases regarding the presence of double peaks. However, its location and amplitude are different. Also we observe a single broad peak in the case of sticky dendrimers. These differences can be due to the ability of dendrimers to stretch and achieve any configuration and size unlike nanoparticles. 

\subsection{Mean squared displacement of dendrimers}\label{sec:MSD}

Theories have been developed to describe nanoparticle dynamics in both permanently crosslinked networks~\cite{cai2015hopping} and associative polymer networks.~\cite{cai2012structure} A detailed description of these scaling theories along with schematic figures can be found in Section~S6 in the Supporting Information. According to these predictions, the dynamics of small particles with diameter $d$ less than the correlation length $\xi$, remain unaffected and follow Stokes–Einstein scaling, while particles with $d > \xi $ exhibit several distinct behaviours~\cite{cai2011mobility,cai2012structure,cai2015hopping}. In a permanently crosslinked, unentangled polymer network, particles larger than the mesh size initially undergo free motion before becoming confined by the surrounding polymer strands, leading to a temporary trapping phase. At longer times, they escape these confinements through rare hopping events, ultimately becoming diffusive. In associative polymer networks, however, the dynamics depend on the particle size relative to the reversible strand size, $r_{st}$, which is the length of the polymer strand between two stickers~\cite{cai2012structure}. For $\xi < d < r_{st}$, the dynamics are similar to those in non-associating polymer liquids, showing an intermediate subdiffusive regime followed by a terminal diffusion. In contrast, for $d > r_{st}$, the motion is strongly influenced by the presence of reversible bonds: after an initial diffusive regime which lasts until $\tau_{\xi}$, which is the relaxation time of a correlation blob, particles undergo subdiffusion until $\tau_{r_{st}}$, where $\tau_{r_{st}}$ is the relaxation time of the polymer strand $r_{st}$ between two neighboring stickers, followed by transient trapping until $\tau_{st}$, where $\tau_{st}$ is the lifetime of a reversible association. A second subdiffusive regime then arises from the sticky-Rouse dynamics of the surrounding chains until $\tau^{st}_d$,  where $\tau^{st}_d$ is the relaxation time of a polymer segment with a size comparable to the particle size, before eventually becoming diffusive at long times.

\begin{figure}[t]
	\begin{center}
                \resizebox{7.5cm}{!} {\includegraphics[trim=80 10 80 0, clip,width=4.0cm]{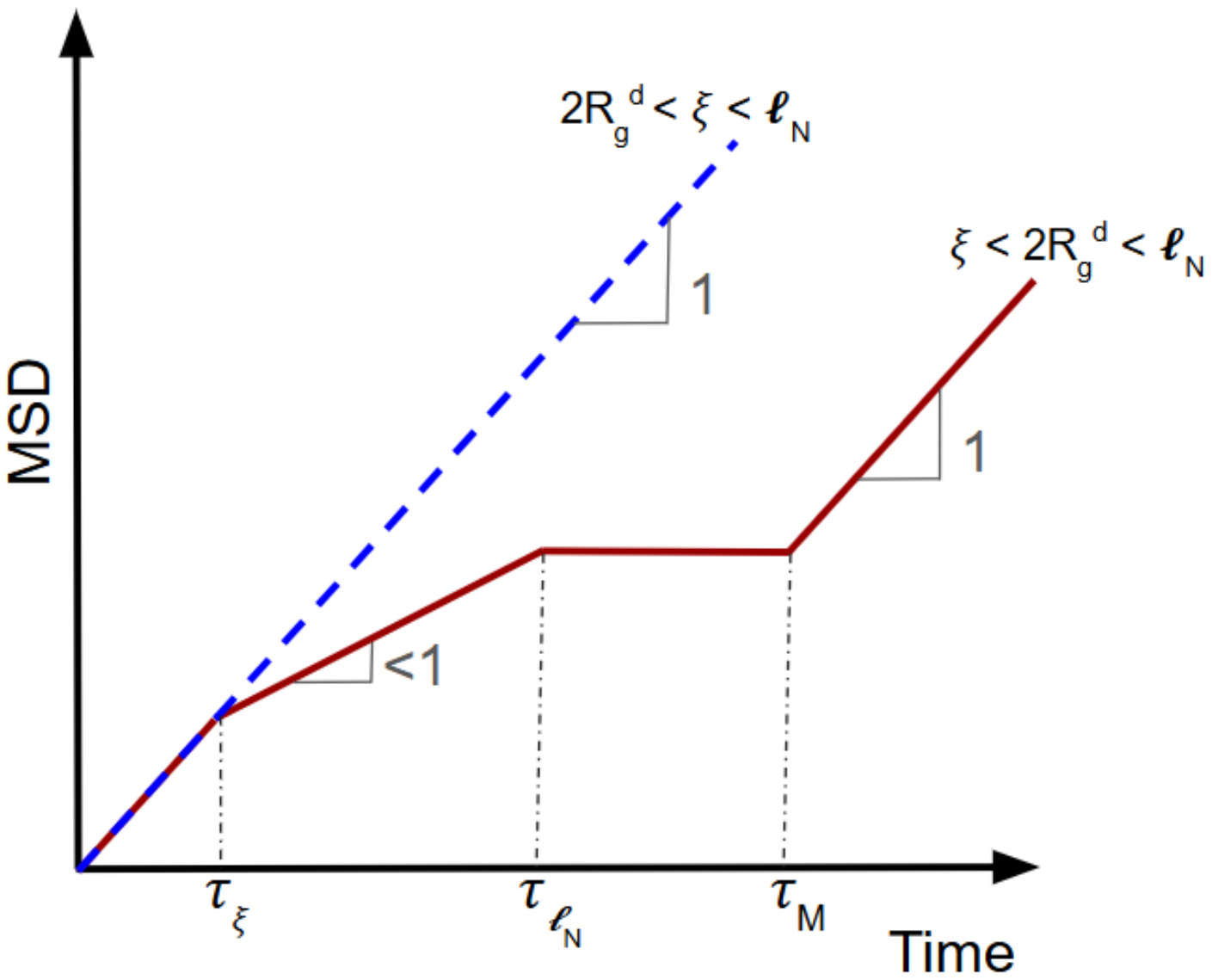}}
	\end{center}
	\vspace{-10pt}
        \caption[Schematic for mean squared displacement of non-sticky dendrimer]{\footnotesize Schematic representation of the mean squared displacement of non-sticky dendrimer belonging to different size regimes with the relevant timescales demarcated.}
    \label{fig:schematic MSD}
\end{figure}
Importantly, the dendrimers in our simulations are observed to be smaller than the strand length $l_{\mathrm{N}}$ and mesh size $a_{\textrm{x}}$ at all concentrations, as shown in Section~S7 of the Supporting Information. This is analogous to the case of $\xi < d < r_{st}$ for nanoparticles in associative polymer networks, where the second subdiffusive regime is absent.
\begin{figure*}[h]
	\begin{center}
		\begin{tabular}{ccc}
			\includegraphics[trim=130 20 130 25, clip,width=0.32\textwidth]{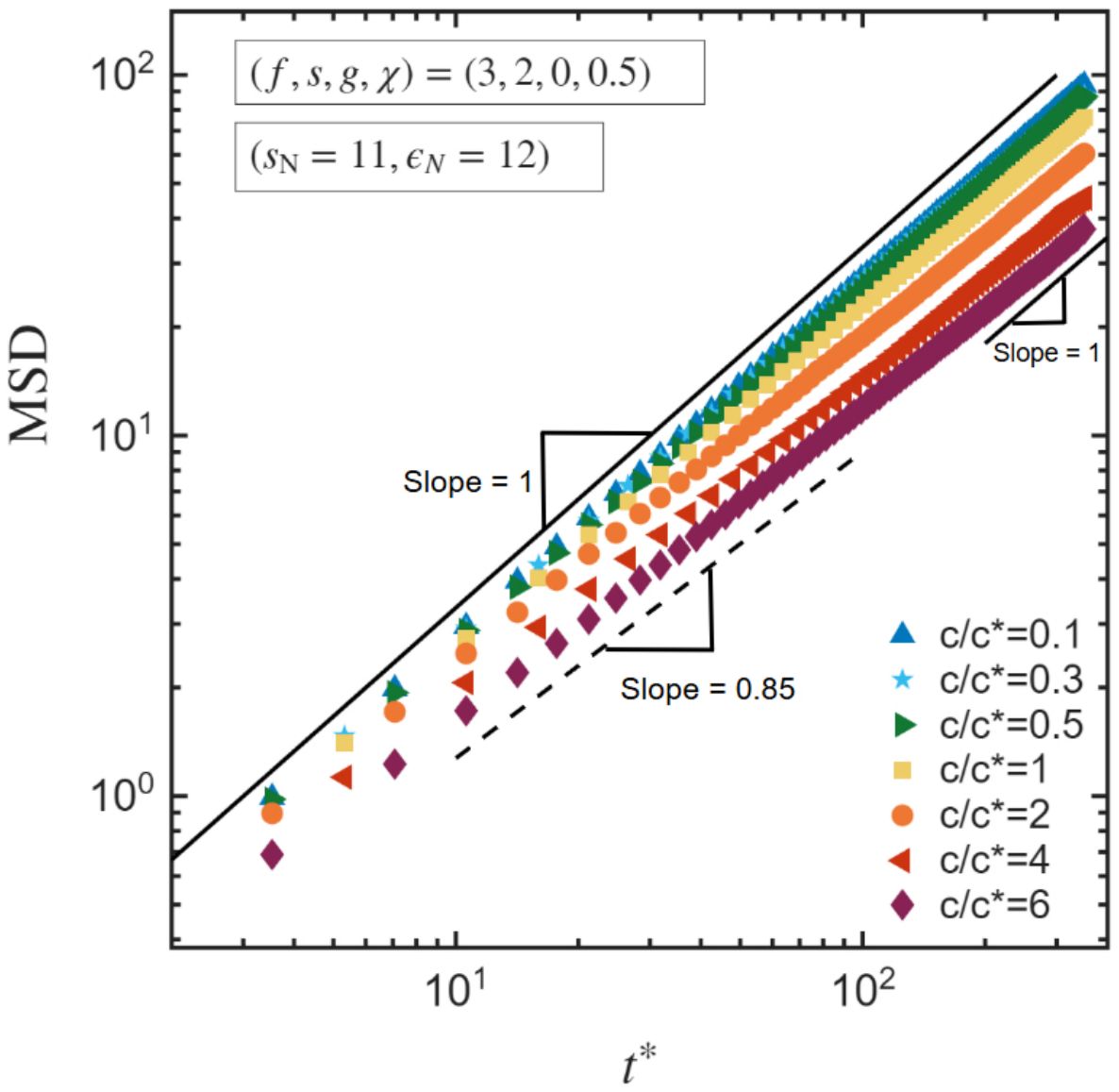} &			
			\includegraphics[trim=130 17 130 25, clip,width=0.32\textwidth]{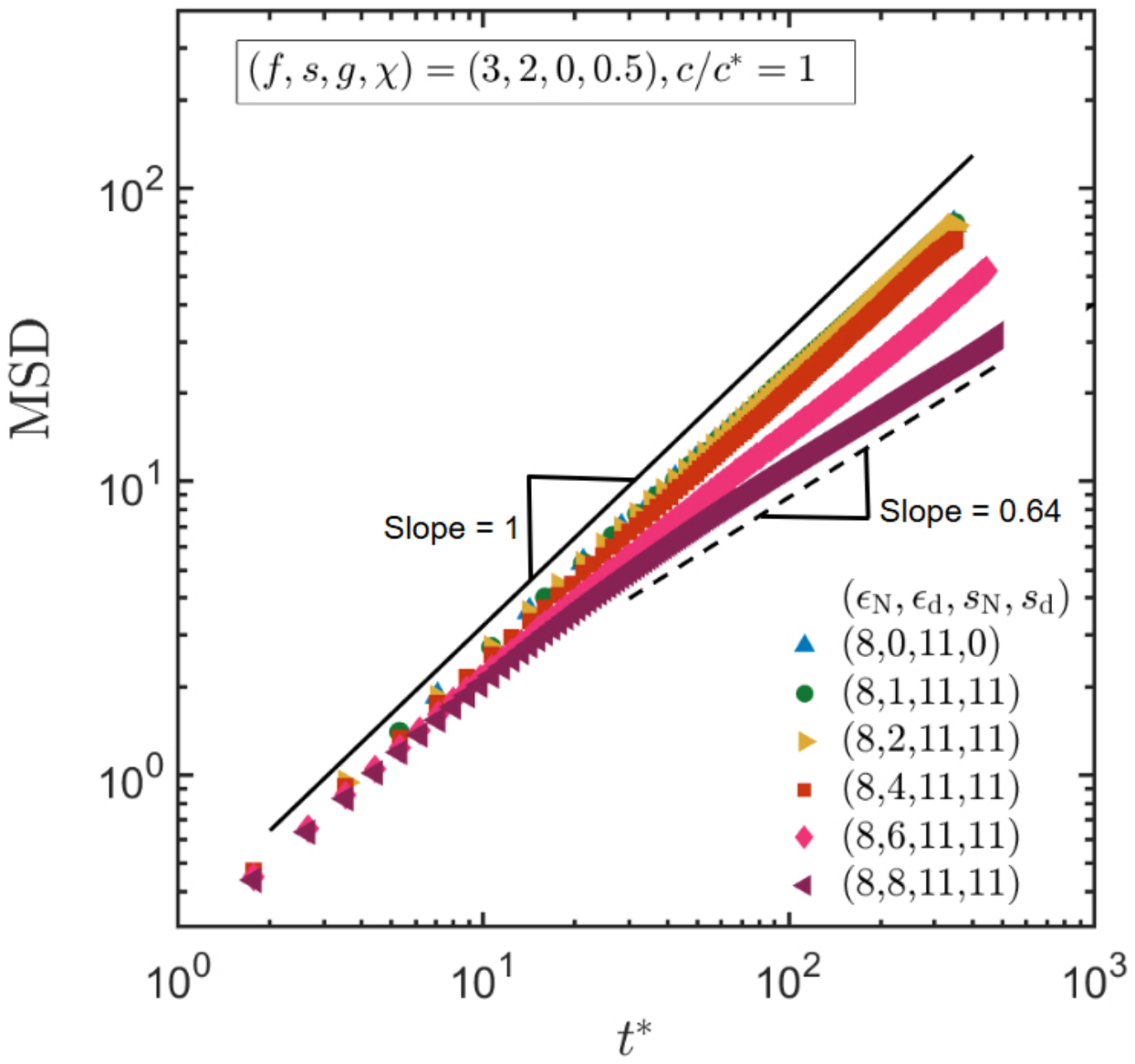} &   
            \includegraphics[trim=10 140 10 80, clip,width=0.32\textwidth]{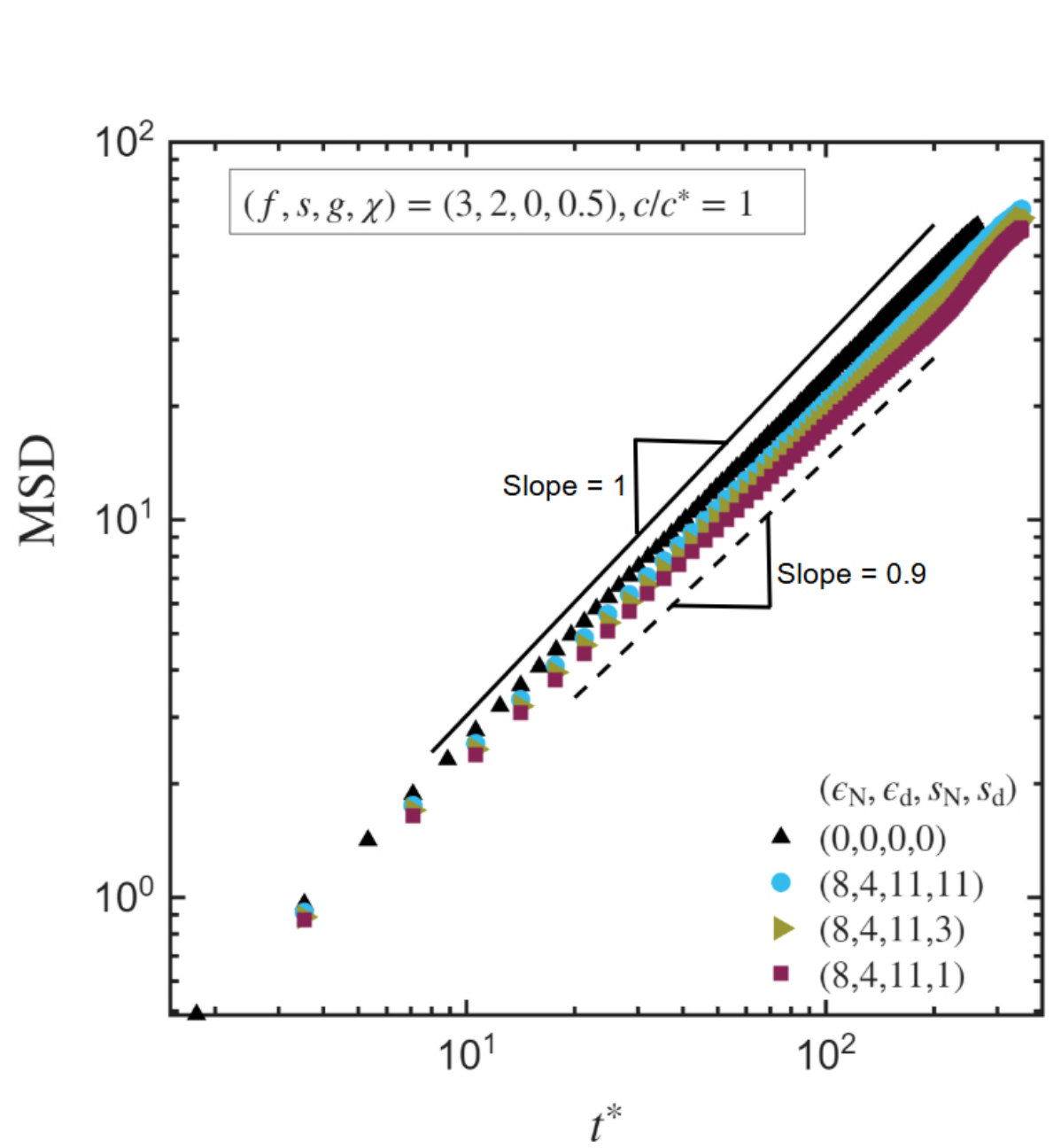} \\
			(a) & (b) & (c)
		\end{tabular}
	\end{center}
	\vspace{-10pt}
        \caption[Effect of dendrimer interaction parameters on sticky dendrimer]{\footnotesize (a) Role of concentration of linear chains on the mean squared displacement of a non-sticky dendrimer. (b) Effect of the sticker strength $\epsilon_{\textrm{d}}$ on the MSD of sticky dendrimers. (c) Effect of the distance between the dendrimer interaction stickers $s_{\mathrm{d}}$ on the MSD
of sticky dendrimers. The straight line in all figures have slope = 1 and the dashed line is a with slope $<1$.}
    \label{fig:msd_sticky}
\end{figure*}

The inclusion of stickers introduces four important timescales that significantly influence the dynamics of sticky dendrimers. These timescales are related to the characteristic distances and interaction strengths in the system, specifically $l_{\textrm{N}}$, $l_{\textrm{d}}$, $\epsilon_{\textrm{d}}$ and $a_{\textrm{x}}$; and are denoted by $\tau_{l_{\textrm{N}}}$, $\tau_{l_{\textrm{d}}}$, $\tau_{\textrm{M}}$ and $\tau_{a_{\textrm{x}}}$ respectively. 

\begin{enumerate}
    \item $\tau_{l_{\textrm{N}}}$: This represents the relaxation time of a chain segment of length equal to $l_{\textrm{N}}$, which corresponds to $s_{\mathrm{N}}$, given by:
    \begin{align}\label{eq:tau_l_N}
    \tau_{l_{\textrm{N}}} \equiv \dfrac{\eta_s l_{\textrm{N}}^3}{k_BT}
    \end{align}
    This time scale is similar to $\tau_{r_{st}}$ defined by Cai \etal ~\cite{cai2012structure} since $r_{st}$ is equivalent to $l_{\textrm{N}}$.
    \item $\tau_{\textrm{M}}$: The time for which a sticker remains bound to another sticker is referred to as $\tau_{\textrm{M}}$, and it increases exponentially with sticker strength~\cite{rubinstein2001dynamics,robe2024evanescent}. It is obtained by calculating an autocorrelation function ($C_{\mathrm{M}} \left( \Delta t \right)$) from the association matrix ($M_2$). The $M_2$ matrix gives information about all the interchain associations of type A stickers. The autocorrelation function is calculated from 
    \begin{align}\label{eq:bond autocorrelation function}
    C_{\mathrm{M}} \left(\Delta t\right) = \langle M_2(t) \cdot M_2(t+\Delta t)\rangle
    \end{align}
    $C_{\mathrm{M}} \left(\Delta t\right)$ is fitted with a single exponential function to obtain the sticker duration, $\tau_{\textrm{M}}$. The algorithm used here to estimate $\tau_{\textrm{M}}$ is similar to that used by Robe \etal.~\cite{robe2024evanescent} Clearly, $\tau_{\textrm{M}}$, which estimates the lifetime of sticker association, is of the order of the timescale $\tau_{st}$ introduced by Cai \etal.~\cite{cai2012structure}    
    \item $\tau_{a_{\textrm{x}}}$: This timescale corresponds to the relaxation time of a chain segment with length $a_{\textrm{x}}$, calculated using:
    \begin{align}\label{eq:tau_a_x}
    \tau_{a_{\textrm{x}}} \equiv \dfrac{\eta_s a_{\textrm{x}}^3}{k_BT}
    \end{align}
    \item $\tau_{l_{\textrm{d}}}$: This corresponds to the relaxation time of a chain segment of length $l_{\textrm{d}}$ (corresponding to $s_{\mathrm{d}}$):
    \begin{align}\label{eq:tau_l_d}
    \tau_{l_{\textrm{d}}} \equiv \dfrac{\eta_s l_{\textrm{d}}^3}{k_BT}
    \end{align} 
    This timescale is new here and is in addition to those predicted by the scaling theories for rigid non-sticky nanoparticles, owing to the attractive interaction between the dendrimers and the background network.
\end{enumerate}

Using the framework of the scaling theory for nanoparticles in physically associating networks, we construct a schematic mean squared displacement plot for non-sticky dendrimers, as shown in Figure~\ref{fig:schematic MSD}. While the theory predicts a step-like transition of the anomalous diffusion exponent from $\alpha=1$ at short times to $\alpha=0.5$ at intermediate times, experiments and simulations for probe particles in various polymeric environments instead reveal a smoother crossover, with $\alpha$ gradually decreasing below unity in the intermediate regime.~\cite{chen2018coupling,kumar2019transport,mariya2024universal,sorichetti2021dynamics}

To test these predictions in the case of soft dendrimers, we calculated the mean squared displacement of the centre of mass of dendrimers in a polymer network using the following expression:
\begin{align}\label{eq:MSD}
    \textrm{MSD}(\Delta t) = \langle | \mathbf R_{\textrm{CM}}(t+\Delta t)-\mathbf R_{\textrm{CM}}(t) |^2 \rangle
\end{align}
where $\mathbf R_{\textrm{CM}}(t)$ and $\mathbf R_{\textrm{CM}}(t+\Delta t)$ are the position vectors of the centre of mass of the molecule at times $t$ and $t+\Delta t$ respectively. The diffusion coefficient $D$ and the diffusion exponent $\alpha$ can be obtained from the mean squared displacement as given below:
\begin{align}\label{eq:MSD_related_to_alpha}
    \textrm{MSD}(\Delta t) = 6 \,D \, t^{\alpha}
\end{align}

As the concentration of background polymer chains increases, the mean squared displacement (MSD) of dendrimers decreases significantly (as shown in Figure~\ref{fig:msd_sticky}(a)), indicating slower diffusion. Interestingly, in contrast to nanoparticles that are expected to exhibit a region of constant MSD due to trapping in polymer cages followed by a hopping mechanism that leads to diffusion at longer times,~\cite{cai2015hopping,sorichetti2021dynamics} dendrimers, whether non-sticky or sticky (as shown in Section~S8,~S9 and ~S10 in the Supporting Information), do not show such long-term confinement at any concentrations or architectures considered in this study. Both types of dendrimers transition from subdiffusive to diffusive regimes at long times without evidence of persistent trapping. This behaviour may be due to the flexible architecture of dendrimers, allowing for shape fluctuations that facilitate their escape from polymer cages, unlike the rigid structure of nanoparticles. 

Consistent with the behaviour of homopolymer dendrimers in semidilute solutions of linear chains, we found that the dendrimer remains diffusive as long as its size is smaller than the correlation length of the solution ($\xi$) (shown in Figure~\ref{fig:msd_sticky}(a)). The concentration at which the onset of subdiffusion is observed depends on the size ratio between the dendrimers and linear chains. For $\chi=0.5$, this happens beyond $1c^{\ast}$, the concentration after which the size of the dendrimer is larger than $\xi$ (refer to Section~S7 in the Supporting Information). 

At a fixed concentration, increasing the sticker strength $\epsilon_{\textrm{d}}$ has a significant impact on the MSD of the sticky dendrimer as shown in Figure~\ref{fig:msd_sticky}(b). At short and long times, sticky dendrimers exhibit normal diffusive motion like nanoparticles~\citep{cai2011mobility,cai2015hopping} and non-sticky dendrimers. However, subdiffusive behaviour ($\alpha < 1$) is observed at intermediate times at low concentrations (where normal diffusion was observed otherwise). For small $\epsilon_{\textrm{d}}$, the dendrimer behaves similar to a non-sticky dendrimer, with almost no effect on its MSD. This behaviour parallels trends observed in the radius of gyration, where weak interactions do not significantly alter dendrimer size. As $\epsilon_{\textrm{d}}$ increases, the number of binding and unbinding events decreases and the dendrimer remains attached to the network for longer durations. This extended residence time results in a slower diffusion rate. The onset of subdiffusion is observed even when the size of the dendrimer is smaller than the network correlation length (which is true at low concentrations), indicating that it is the dendrimer-linear chain interactions that primarily influence sticky dendrimer dynamics. 

\begin{figure*}[t]
	\begin{center}
		\begin{tabular}{ccc}
			\resizebox{7.5cm}{!} {\includegraphics[trim=100 10 130 15, clip,width=4.0cm]{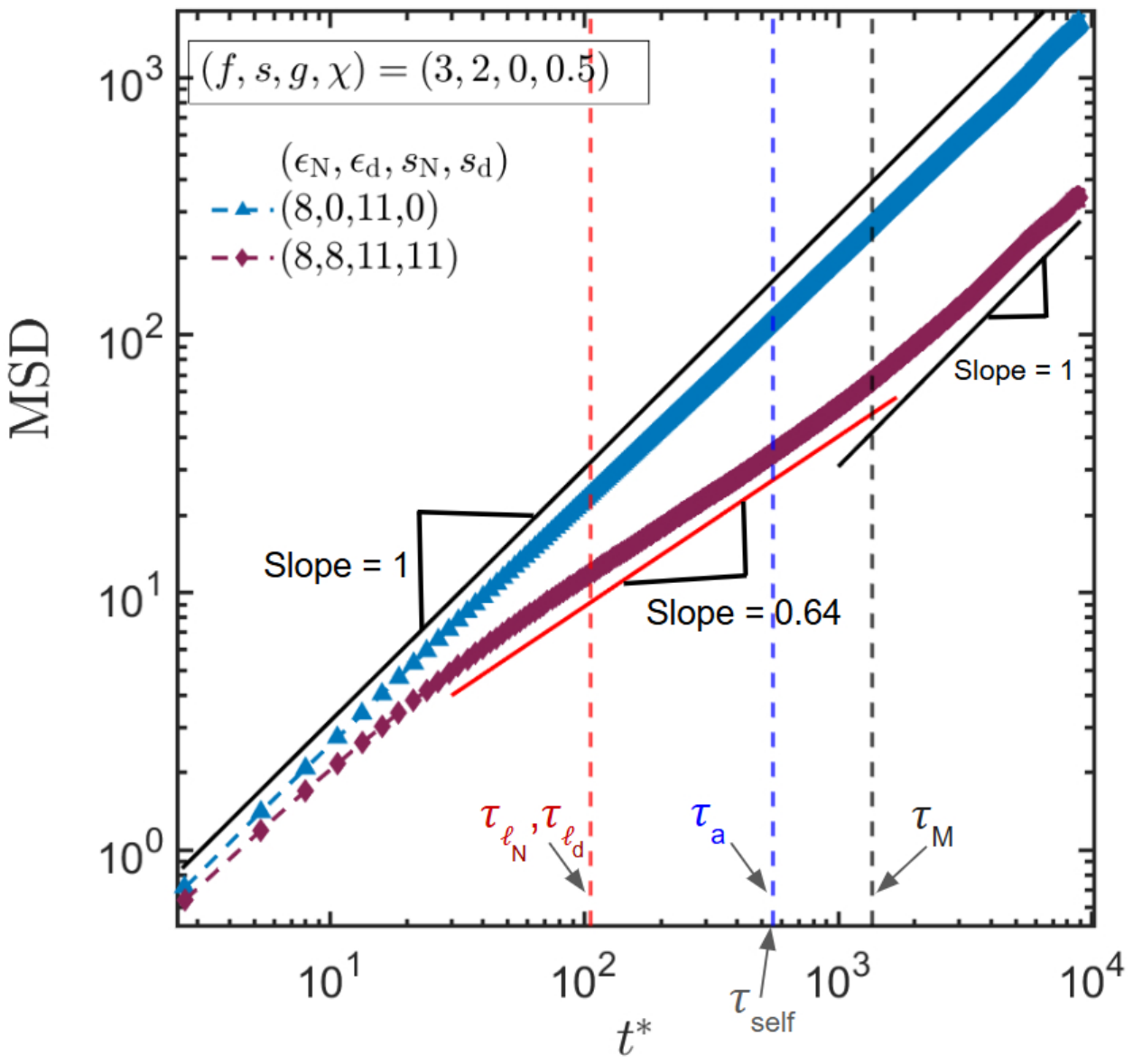}} &   
                \resizebox{7.5cm}{!} {\includegraphics[trim=100 10 130 15, clip,width=4.0cm]{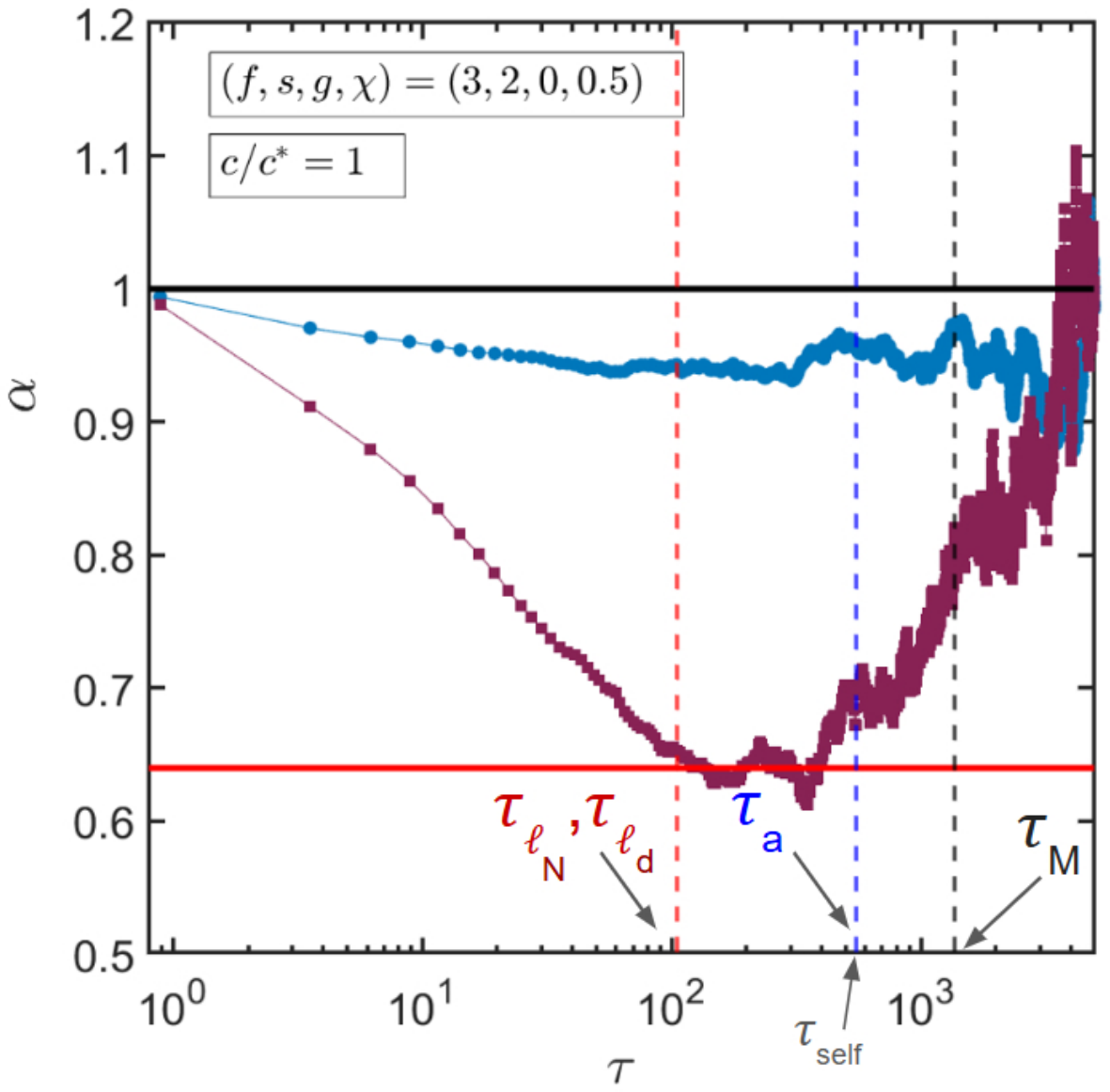}}\\
			(a) & (b)
		\end{tabular}
	\end{center}
	\vspace{-10pt}
        \caption[MSD and diffusion exponent as functions of time]{\footnotesize (a) The mean squared displacement of sticky and non-sticky dendrimers of identical architectures in polymer network at $c/c^{\ast}=1$. (b) The diffusion exponent as a function of time for sticky and non-sticky dendrimers. The system considered is the same as that in (a). The black horizontal line indicates $\alpha=1$ and the red horizontal line is $\alpha=0.64$ which is the slope of the red line in the subdiffusive regime in (a). The dashed lines in both figures show the important timescales involved.}
    \label{fig:MSD and alpha as a function of time}
\end{figure*}

To understand the effect of the number of stickers, we varied $s_{\mathrm{d}}$ at a constant $\epsilon_{\textrm{d}}$. Reducing $s_{\mathrm{d}}$ increases the number of potential binding sites available to the dendrimer. This leads to a decrease in the MSD, though the effect is less pronounced compared to changes in sticker strength (shown in Figure~\ref{fig:msd_sticky}(c)). The effect of $\epsilon_{\textrm{d}}$ and $s_{\mathrm{d}}$ on sticky dendrimers is similar across different dendrimer architectures as shown in Section~S10 of the Supporting Information. Note that the network-related parameters like the sticker strength of the network forming stickers ($\epsilon_{\textrm{N}}$) and the distance between them ($s_{\mathrm{N}}$) do not affect the dynamics of sticky dendrimers as shown in Figures S8 and S9 in the Supporting Information.

\subsection{Comparison between dynamics of sticky and non-sticky dendrimers}

The influence of interactions between the dendrimer and the linear chain network is investigated by comparing the dynamics of sticky and non-sticky dendrimers with sizes smaller than the network mesh size or correlation length. Figure~\ref{fig:MSD and alpha as a function of time}(a) shows the mean squared displacement as a function of time for both types of dendrimers estimated using Eq.~\ref{eq:MSD}. It is evident from this plot that the adhesion of the dendrimer to the network causes it to become subdiffusive. However, the observed subdiffusion is restricted to intermediate times, with the dendrimers being diffusive at short and large times. 

The initial diffusive regime is very short for sticky dendrimers, which then quickly transitions to subdiffusion. The slope of the mean squared displacement plot attains a minimum at $\tau_{l_{\textrm{d}}}$, which is of the order of the time required for the dendrimer to encounter stickers on the network. Due to the combined effect of sticking and trapping, the sticky dendrimer remains subdiffusive. To transition back to the diffusive regime, the dendrimer must either unbind from the network or be released through the fluctuations in the cage-forming strands. Consequently, this return to the diffusive regime typically happens at times much larger than $\tau_{a_{\textrm{x}}}$ and $\tau_{\textrm{M}}$ as shown in Figure~\ref{fig:MSD and alpha as a function of time}(a). Another important timescale we estimated is the self-diffusion time, $\tau_{\mathrm{self}}$, of the sticky dendrimer (($\epsilon_{\mathrm{N}},\epsilon_{\mathrm{d}},s_{\mathrm{N}},s_{\mathrm{d}})=(8,8,11,11)$) calculated using $\tau_{\mathrm{self}} = (2 R_g^d)^2 / (6D)$, where $R_g^d$ is the radius of gyration of the dendrimer and $D$ is the long time diffusion coefficient obtained from the linear terminal regime of the MSD. We find that the binding timescale $\tau_{l_{\textrm{d}}}$ is smaller than $\tau_{\mathrm{self}}$, indicating that the dendrimer binds to the network before diffusing over its own size. Consistently, the ratio of the dendrimer size to the radius of gyration of a chain containing $s_{\mathrm{d}}$ beads is greater than 1 ($2R_g^d/R_g^{s_d}=1.75$), showing that it encounters multiple stickers within a distance shorter than its own dimension, explaining the early onset of subdiffusive dynamics.

To quantify the nature of dendrimer motion in a network further, we examine the diffusion exponent $\alpha$, which characterises the scaling of the mean-squared displacement with time and distinguishes between normal and anomalous diffusion regimes. It is calculated by computing the instantaneous derivative of the MSD as given below:
\begin{align}\label{eq:alpha_time}
    \alpha= \frac{d \log (\textrm{MSD}(\tau))}{d \log \tau}
\end{align}

\begin{figure*}[!h]
	\begin{center}
		\begin{tabular}{cc}
			\resizebox{7.0cm}{!} {\includegraphics[trim=125 10 125 15, clip,width=4.0cm]{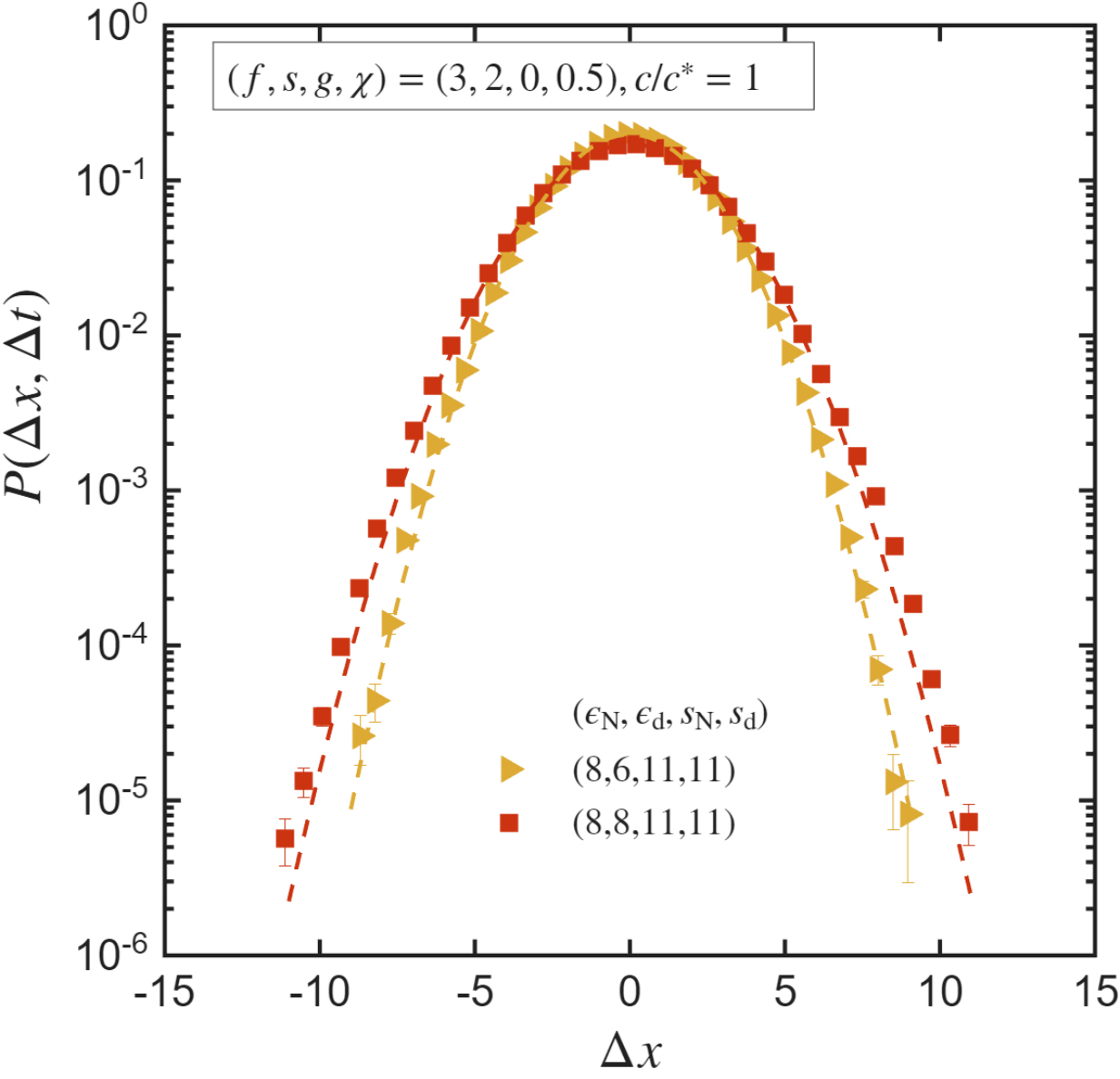}} &
			\resizebox{7.0cm}{!} {\includegraphics[trim=125 10 125 15, clip,width=4.0cm]{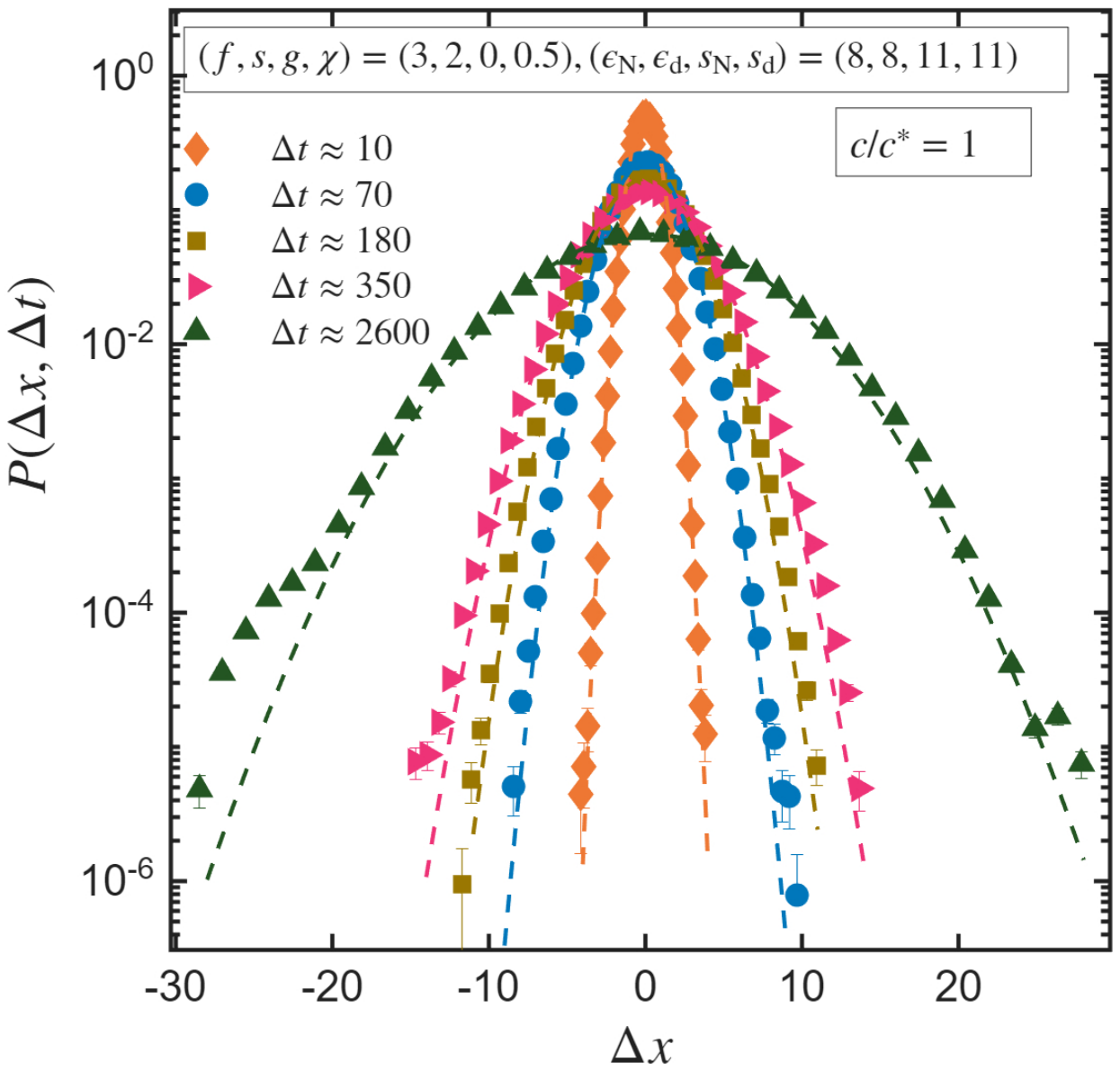}} \\
                (a) & (b) \\
                \resizebox{7.0cm}{!} {\includegraphics[trim=130 10 130 15, clip,width=4.0cm]{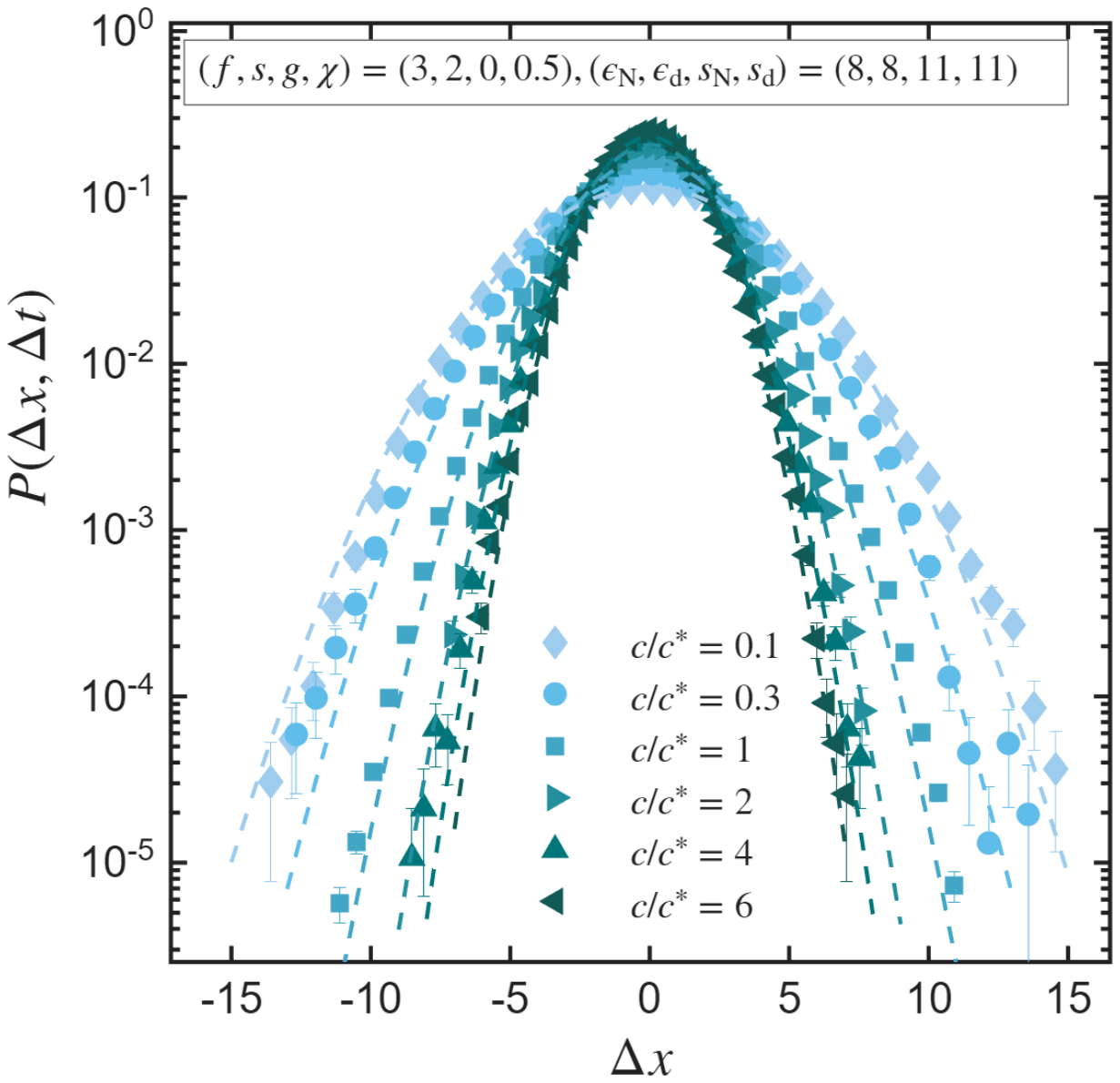}} &
                \resizebox{7.0cm}{!} {\includegraphics[trim=130 10 130 15, clip,width=4.0cm]{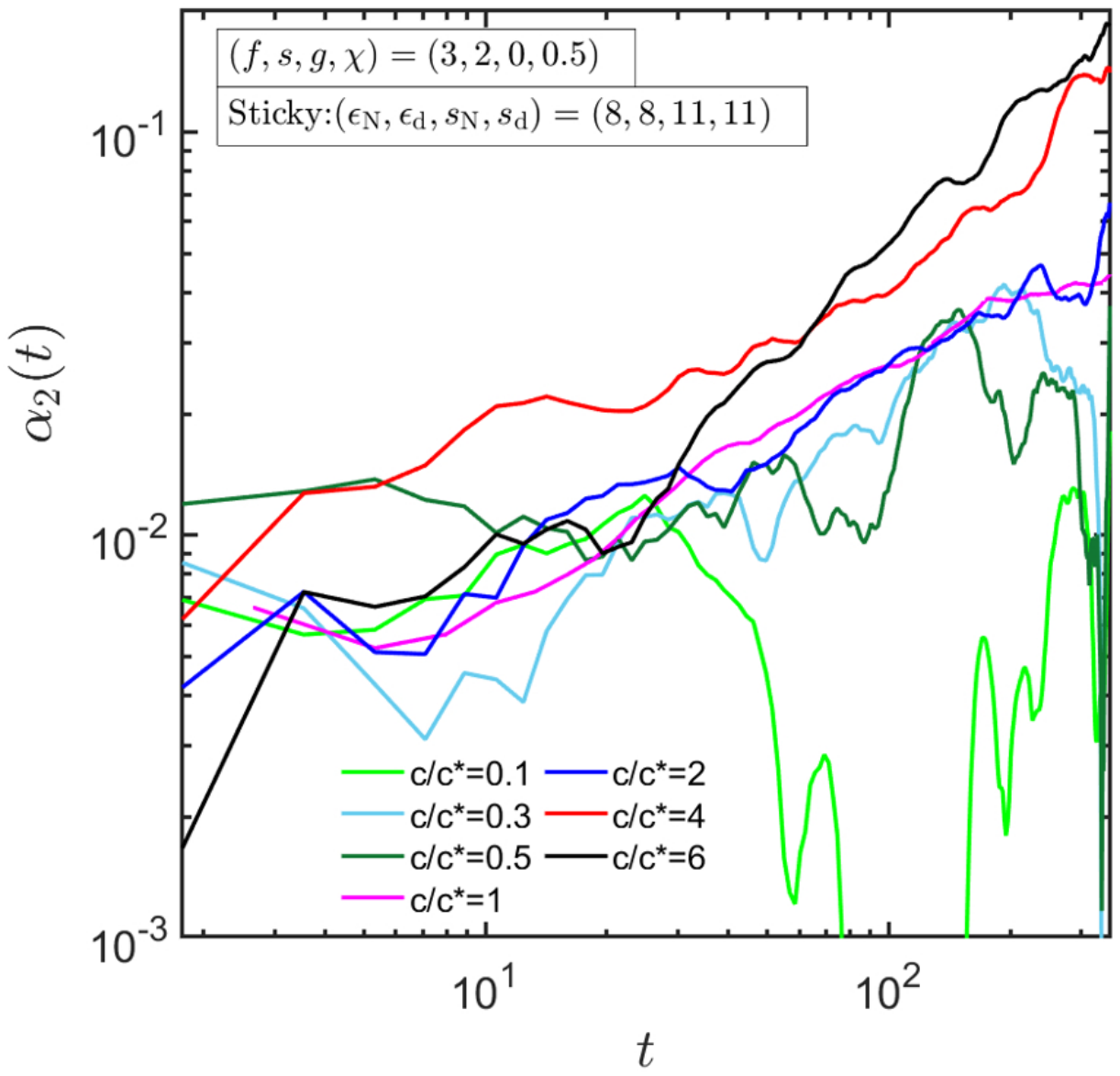}} \\
			 (c) & (d)
		\end{tabular}
	\end{center}
	\vspace{-10pt}
        \caption[Probability distribution function of displacement for sticky dendrimers]{\footnotesize (a) Effect of the sticker strength of the network-dendrimer interaction stickers on the probability distribution function of displacement. The times at which PDF is calculated correspond to the midpoint of the subdiffusive regime. (b) The probability distribution functions are calculated at different times. (c) Effect of concentration of linear chains on the probability distribution function of displacement of a sticky dendrimer. The dashed lines in all figures are Gaussian fits. The system considered in (b) and (c) is ($\epsilon_{\textrm{N}}, \epsilon_{\textrm{d}}, s_{\mathrm{N}}, s_{\mathrm{d}}$)=($8,8,11,11$). (d) The non-Gaussian parameter, $\alpha_2(t)$ (given by Eq~\eqref{eq:NGP}), as a function of time at different concentrations for the sticky dendrimers considered in (c). Tails of the distributions were also analyzed by plotting $\ln(-\ln P(\Delta x,\Delta t))$ versus $\ln \Delta x$, where a Gaussian corresponds to slope $\beta=2$. The fitted slopes are consistently less than 2, confirming slight deviations from Gaussian distributions (see Section~S12 in the SI).}
    \label{fig:PDD of sticky dendrimer}
\end{figure*}

The non-sticky dendrimer exhibits diffusive behaviour across all observed timescales, with a diffusion exponent of approximately $\alpha \approx 1$, as shown in Figure~\ref{fig:MSD and alpha as a function of time}(b). In contrast, the diffusion exponent of the sticky dendrimer decreases to values below unity at intermediate times, indicating subdiffusive dynamics, before recovering to $\alpha \approx 1$ at long times. All the key timescales are marked by the vertical dashed lines. At short times, the dendrimer does not significantly interact with the network strands or sticker sites, resulting in purely diffusive motion. As time progresses, its mobility becomes increasingly hindered due to transient binding to the network, leading to a minimum in the diffusion exponent beyond $\tau_{l_{\textrm{d}}}$. Because the sticker interactions are transient, the dendrimer eventually escapes these constraints on timescales of the order of $\tau_{\textrm{M}}$. Horizontal lines serve as visual guides corresponding to $\alpha=1$ and $\alpha=0.64$, the latter derived from power-law fits to the intermediate-time MSD data shown in Figure~\ref{fig:MSD and alpha as a function of time}(a). As highlighted in the figure, our simulations do not exhibit $\alpha=0.5$ predicted by Cai \etal ~\cite{cai2015hopping}, but instead show a gradual crossover where $\alpha$ decreases gradually below unity in the intermediate regime and later to unity at long times.

\vspace{-10pt}
\subsection{Probability distribution function of displacement}

Probe particles that exhibit subdiffusion are often characterised by non-Gaussian probability distribution of displacements, as a result of the hopping movement of the tracers and heterogenities of the environment.~\cite{phillies2015complex,kumar2019transport,wang2012brownian,smith2021dynamics,he2016dynamic,jee2014nanoparticle} However, in the case of a non-sticky dendrimer diffusing in a solution of linear chains, our previous study revealed that the probability distribution of displacement remained Gaussian regardless of the time at which it was calculated and the concentration of the solution, even though subdiffusion was observed.~\citep{mariya2024universal} This suggests that non-sticky dendrimer dynamics are consistent with fractional Brownian motion, characterised by temporally correlated, yet continuous and ergodic trajectories.~\citep{mandelbrot1968fractional} The probability distribution function of displacement of a molecule displaced by a distance $\Delta x$ in a time $\Delta t$ is given by
\begin{align}\label{eq:PDD}
    \textrm{P}(\Delta x,\Delta t) = \langle \delta \left(\Delta x -  |x_{\textrm{CM}}(t+\Delta t)- x_{\textrm{CM}}(t)|\right) \rangle
\end{align}
where $x_{\textrm{CM}}(t)$ and $ x_{\textrm{CM}}(t+\Delta t)$ are the x-components of the centre of mass of the molecule at times $t$ and $t+\Delta t$ respectively. 

In the case of sticky dendrimers, which also exhibit subdiffusion, we observe only slight deviations from Gaussian displacement distributions, most evident at intermediate times (Figure~\ref{fig:PDD of sticky dendrimer}(a)). These deviations become more noticeable at higher sticker strengths ($\epsilon_{\mathrm{d}}$), where stronger binding to the polymer network enhances confinement. At lower sticker strengths, the distributions remain closer to Gaussian, indicating that weak interactions do not substantially alter dendrimer displacements. At short times, the diffusion of sticky dendrimers is essentially Gaussian, consistent with nearly free motion. As the system enters the subdiffusive regime, however, small but systematic departures from Gaussian behavior emerge (Figure~\ref{fig:PDD of sticky dendrimer}(b)), reflecting the influence of transient binding to the polymer network. Similar marginal deviations from Gaussianity are also observed at intermediate times when varying the polymer concentration (Figure~\ref{fig:PDD of sticky dendrimer}(c)), again pointing to the role of confinement and binding in shaping the displacement statistics. These weak but consistent deviations from Gaussianity are further reflected in the non-Gaussian parameter, $\alpha_2(t)$ given by
\begin{align}\label{eq:NGP}
    \alpha_2(t) = \dfrac{3 \langle r^4(t)\rangle}{5 \langle r^2(t)\rangle^2}-1
\end{align}

\noindent For a Gaussian distribution of displacements, the non-Gaussian parameter $\alpha_2(t)$ is zero, while $\alpha_2(t) > 0$ indicates deviations from Gaussianity. Figure~\ref{fig:PDD of sticky dendrimer}(d) presents the time evolution of $\alpha_2(t)$ in the subdiffusive regime for the sticky dendrimer system shown in Figure~\ref{fig:PDD of sticky dendrimer}(c). The values of $\alpha_2(t)$ remain small ($<0.1$) but consistently non-zero, indicating marginal deviations from Gaussian behavior. While the concentration dependence is relatively weak and somewhat noisy, at longer times there is a tendency for higher concentrations to display larger $\alpha_2$ consistent with increased confinement within the network. This suggests that dendrimer motion is not purely Gaussian, reflecting intermittent mobility due to transient caging and attractive interactions with the network. Occasional detachment events allow relatively freer motion before reattachment, leading to slightly heavier-tailed displacement statistics. Such features are qualitatively consistent with ideas from heterogeneous transport models, such as continuous-time random walks (CTRW), although the present data do not allow us to make a definitive claim. Further, we also analyzed the tails of the distributions by plotting  $\ln (-\ln P(\Delta x, \Delta t))$ versus $\ln \Delta x$. In this representation, $P(\Delta x, \Delta t) \sim \exp(-\frac{x^{\beta}}{2\sigma^2})$ appears as a straight line with slope $\beta$ at large displacements. The fitted slopes are consistently less than 2, confirming the presence of heavier-than-Gaussian tails which is discussed in Section~S12 in the Supporting Information. It is also worth noting that the distance between the dendrimer interaction stickers ($s_{\mathrm{d}}$) does not seem to affect the probability distribution function of sticky dendrimers, as shown in Section~S11 in the SI.

\subsection{Terminal diffusivity}\label{sec:terminal diffusivity}

Theoretical frameworks describing the diffusion of nanoparticles within crosslinked polymer networks propose relations between the terminal diffusivity and the confinement parameter. The hopping theory introduced by by Cai \etal ~\cite{cai2015hopping} provides a scaling relation for the long-time diffusivity of nanoparticles in unentangled polymer networks, expressed as follows:
\begin{equation}\label{eq:scaling law for diffusivity}
    D_{\textrm{hop}} \approx (\xi^2/\tau_x) \frac{\exp(-C^2)}{C}
\end{equation}
where $\tau_x$ is the Rouse relaxation time of a network strand $a_{\textrm{x}}$ and $C$ is the confinement parameter. 

The subsequent theoretical framework proposed by Dell and Schweizer~\citep{dell2014theory} is based on the nonlinear Langevin equation and polymer reference interaction site model (PRISM) theory.~\citep{schweizer1997integral,hall2008many} Their predictions indicate that the onset of hopping occurs beyond a critical value of the confinement parameter, denoted as $C_c$. Beyond this threshold, the nanoparticle becomes confined within network cages, and hopping is the only process for nanoparticle mobility. The hopping diffusion coefficient, $D_{\textrm{N}}$, can be estimated by the relation 
\begin{equation}\label{eq:Dell_shweizer}
    D_{\textrm{N}} \approx \Delta_h^2/\tau_h
\end{equation} 
where $\Delta_h$ represents the mean jump length and $\tau_h$ is the mean hopping time. 

Recently, using MD simulations, Sorichetti \etal~\cite{sorichetti2021dynamics} have demonstrated that both these theoretical frameworks hold true in the case of repulsive and weakly attractive nanoparticles in a crosslinked polymer network. Based on Eq.~\ref{eq:scaling law for diffusivity} from the hopping theory, they used the following relation to fit their simulation data for the terminal diffusivity $D_{\textrm{N}}$:
\begin{equation}\label{eq:Cai scaling Kob}
    D_{\textrm{N}} = A \exp [ -\left( B C\right)^{2}] / BC
\end{equation}
where $A$ and $B$ are fitting parameters, thus validating the hopping theory. By incorporating the dependencies of the mean jump length $\Delta_h$ on the confinement parameter and $\tau_h$ on free energy barrier in Eq.~\ref{eq:Dell_shweizer}, Sorichetti \etal~\cite{sorichetti2021dynamics} propose that $D_{\textrm{N}}$ also depends on $C$ as follows:
\begin{equation}\label{eq:Dell_shweizer kob}
    D_{\textrm{N}} = A' \exp [ -\left( B' C\right)^{\delta}]
\end{equation}
where $A',B'$ and $\delta$ are fitting parameters. Using an appropriate choice of the fitting parameters, Sorichetti \etal~\cite{sorichetti2021dynamics} show that their simulations follow the analytical predictions of Dell and Schweizer~\cite{dell2014theory} as well. Eq.~\ref{eq:Dell_shweizer kob} seems to be structurally similar to Eq.~\ref{eq:Cai scaling Kob}, however, the two are different due to the latter's dependence on the confinement parameter $C$, at small $C$ (in the denominator).

\begin{figure}[t]
	\begin{center}
                \resizebox{8.4cm}{!} {\includegraphics[trim=125 10 125 15, clip,width=4.0cm]{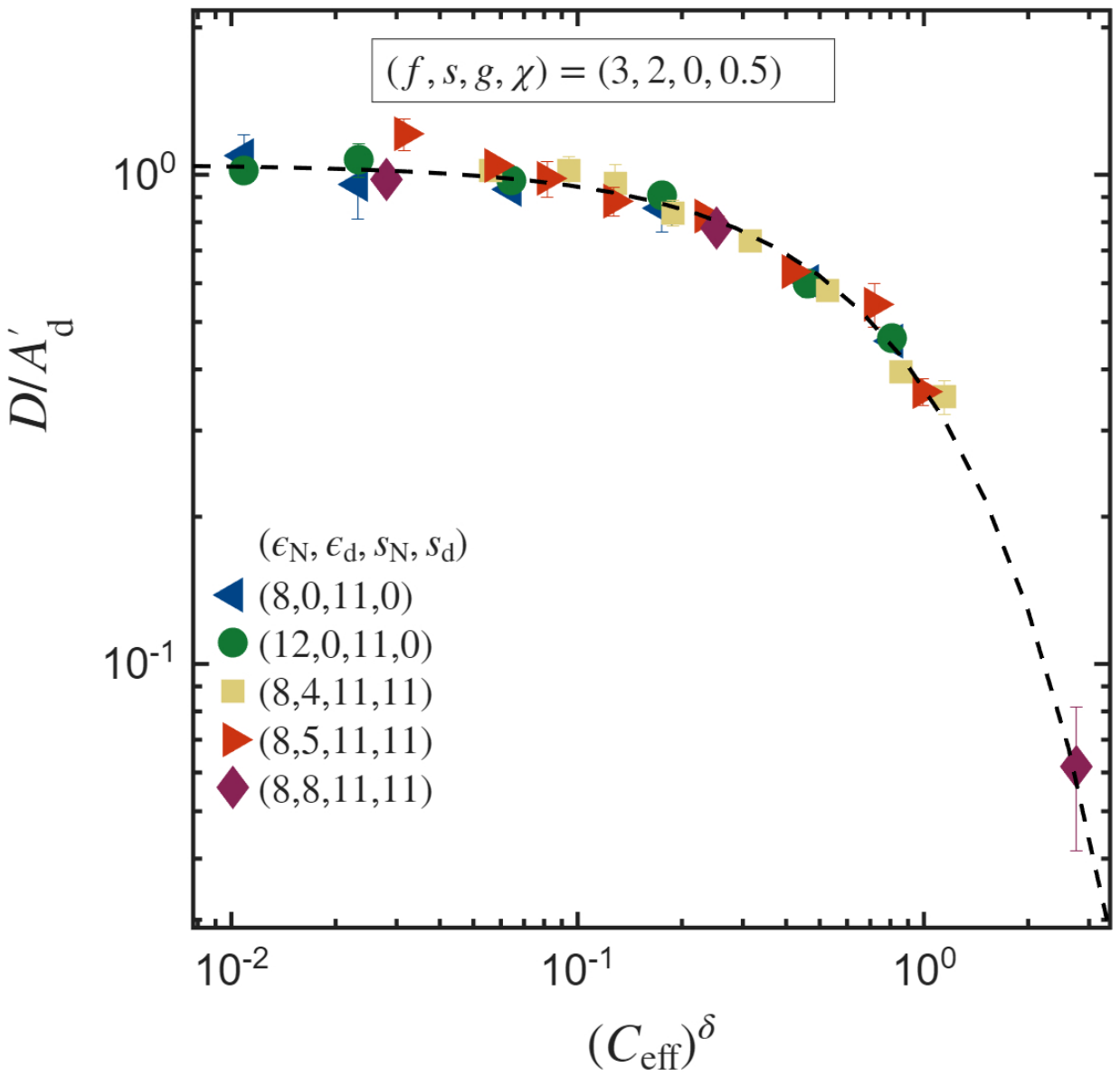}}
	\end{center}
	\vspace{-10pt}
        \caption[Terminal diffusivity of dendrimers]{\footnotesize The normalised terminal diffusivity of non-sticky and sticky dendrimers as a function of normalised confinement parameter. The dashed line is Eq.~\ref{eq:Dell_shweizer kob}.}
    \label{fig:terminal diffusivity}
\end{figure}

\begin{table}[!h]
    \caption{\footnotesize Values of fitting parameters used in Figure~\ref{fig:terminal diffusivity}. Various systems are reported using a combination of ($\epsilon_{\textrm{N}},\epsilon_{\textrm{d}},s_{\mathrm{N}},s_{\mathrm{d}}$).}
    \label{tab:fitting_params}
    \centering
    \setlength{\tabcolsep}{10pt}
    \renewcommand{\arraystretch}{1.5}
    \begin{tabular}{|c|c|c|c|}
         \hline
 $\textrm{System}$     & $A'_{\textrm{d}}$ 
            & $B'_{\textrm{d}}$  & $\delta_{\textrm{d}}$
            \\                
\hline
\hline
            $(8,0,11,0)$
            & $1.01$
            & $-0.25$
            & $2.00$
\\
            $(12,0,11,0)$
            & $1.01$
            & $-0.25$
            & $2.00$
\\
            $(8,4,11,11)$
            & $0.04$
            & $0.31$
            & $1.00$
\\
            $(8,5,11,11)$
            & $0.03$
            & $0.27$
            & $1.14$
\\
            $(8,8,11,11)$
            & $1.77$
            & $1.39$
            & $2.61$
\\
\hline
    \end{tabular}
    \end{table}

The normalised diffusivity of a number of gel-solute systems has also been studied based on the free volume and obstruction theories.~\citep{amsden1998solute,masaro1999physical,axpe2019multiscale} However, these approaches do not account for the effect of the flexibility of the network polymer chains, which has been recently identified as a crucial factor influencing nanoparticle dynamics.~\citep{kumar2019transport,godec2014collective} Therefore, Quesada-Pérez \etal~\cite{quesada2021universal} proposed a unified model for the relative diffusivity of nanoparticles in flexible, crosslinked hydrogels, expressed as a function of the parameter $\beta$: 
\begin{equation}\label{eq:quesada equation}
    D/D_0=\exp \left( -1.77 \beta^{1.07} \right)
\end{equation}
where $\beta = \phi \left( 1 + R_s/R_p\right)$, $\phi$ is the volume fraction of polymer solution, $R_s$ is the radius of the solute and $R_p$ is the radius of the polymer. A recent simulation study has validated this theory for solutes of varying sizes in polymer networks with different bond stiffness.~\citep{quesada2022coarse} However, these simulations did not account for hydrodynamic interactions. 

In our previous study,~\cite{mariya2024universal} it has been shown that dendrimers, owing to their size fluctuations, do not follow the scaling law developed for terminal diffusivity of nanoparticles in semidilute polymer solutions. Instead, they have a different scaling exponent, which changes with increasing functionality. To investigate if the nanoparticle theories in polymer networks apply to floppy dendrimers, we calculated the terminal diffusivity of sticky and non-sticky dendrimers in the longtime diffusive regime. The confinement parameter for dendrimers is given below:
\begin{align}\label{eq:confinement parameter}
    C_{\textrm{d}}= \frac{2 R_{\textrm{g}}^{\textrm{d}}}{\xi}
\end{align}
Following the arguments of Sorichetti \etal,~\cite{sorichetti2021dynamics}, we study the dependence of normalised diffusivity $D/A'_{\textrm{d}}$ on the normalised confinement parameter $\left(C_{\textrm{eff}}\right)^{\delta_{\textrm{d}}}$, where $C_{\textrm{eff}}=B'_{\textrm{d}}C_{\textrm{d}}$. Both the non-sticky and sticky dendrimers, having a range of sticker energies, follow Eq.~\ref{eq:Dell_shweizer kob} as shown in Figure~\ref{fig:terminal diffusivity}. The values of the various coefficients are given in Table~\ref{tab:fitting_params}. These are different from those reported by Sorichetti \etal,~\cite{sorichetti2021dynamics} for nanoparticles. There are deviations from Eq.~\ref{eq:Dell_shweizer} at lower $\left(C_{\textrm{eff}}\right)^{\delta_{\textrm{d}}}$, which corresponds to small $C_{\textrm{d}}$. This is expected as the theory is derived for higher confinement parameters. The fitting parameters for the two non-sticky dendrimer cases considered are the same since they exhibit similar diffusive dynamics as discussed in Section~S8 in the Supporting Information. However, sticky dendrimers show decreased mobility when the sticker strength $\epsilon_{\textrm{d}}$ is increased, hence they require different fitting parameters. Additionally, we have reported the terminal diffusivities of systems only for which we were able to acquire data in the terminal diffusive regime in the case of sticky dendrimers. Note that dendrimers in our simulations do not follow Eq.~\ref{eq:scaling law for diffusivity} or Eq.~\ref{eq:quesada equation}. 

\section{\label{sec:conclusions}Conclusions}

The transport of various types of probe particles through polymer solutions and networks holds significant interest for  biomedical,~\citep{siepmann2012modeling,peppas2000hydrogels} industrial~\citep{hermans1968role} and biological applications.~\citep{chatterjee2011subdiffusion,tabei2013intracellular,goodrich2018enhanced} Some of these particles interact with their surroundings while others undergo simple Brownian motion. This study used Brownian dynamics simulations to examine dendrimers as probe particles within networks of associative polymers, focusing on how both non-sticky and sticky interactions affect network structure and dendrimer behaviour. This is achieved by having various types of stickers in the system, having specific functions (type A for network formation, type B and C for network-dendrimer interactions present on the respective polymer architectures). Essential interactions like excluded volume and hydrodynamic interactions are included. 

We find that the size of non-sticky dendrimers and sticky dendrimers with low sticker interaction strengths remains constant in the dilute regime and decreases in the semidilute regime, similar to the behaviour of non-sticky dendrimers in polymer solutions. In contrast, sticky dendrimers with high interaction strength ($\epsilon_{\textrm{d}}$) exhibit a non-monotonic size behaviour: they initially swell at low concentrations but shrink at higher concentrations, collapsing onto a master scaling curve. Notably, the spacing between stickers on the linear chains ($s_{\mathrm{d}}$) does not influence the dendrimer size. This trend is supported by the bead density distribution profiles, which show that the density of beads away from the dendrimer core initially increases with concentration before declining. Regardless of stickiness or concentration, all dendrimers maintain a dense core with bead density gradually decreasing toward the periphery. Analysis of the network's mesh size distribution reveals that non-sticky dendrimers induce the formation of localized cavities, resulting in a bimodal distribution. In contrast, attractive interactions between sticky dendrimers and the network stretch the mesh, leading to a broader distribution of mesh sizes.

The mean squared displacement (MSD) of non-sticky dendrimers in a polymer network exhibits key features characteristic of their behaviour in semidilute polymer solutions. When the dendrimer size is smaller than the network correlation length ($\xi$), their motion remains diffusive at all times. At higher concentrations, where the dendrimer size exceeds $\xi$, a subdiffusive regime emerges at intermediate times, preceded by short-time diffusion and followed by long-time diffusive behaviour. Notably, the strength and spacing of the network-forming stickers do not significantly affect the dynamics of non-sticky dendrimers. In contrast, sticky dendrimers exhibit subdiffusive motion even when their size is smaller than $\xi$. Their MSD decreases with increasing interaction strength ($\epsilon_{\textrm{d}}$) and with decreasing spacing between type B stickers on the linear chains. This behaviour is attributed to the increased residence time of sticky dendrimers within the network due to attractive interactions. Furthermore, we observe a slightly non-Gaussian displacement distribution for sticky dendrimers in the subdiffusive regime, consistent with previous observations for nanoparticles and flexible viruses. The deviations are weak but systematic, and their extent increases with both $\epsilon_{\textrm{d}}$ and polymer concentration. Finally, we note that the terminal diffusivities of both sticky and non-sticky dendrimers obtained from our simulations are in agreement with the theoretical predictions for nanoparticles in comparable polymer network environments. Thus our study provides fundamental insight into how transient binding and network confinement together governs the dynamics of soft colloidal probes in polymer networks. These findings are broadly relevant for understanding transport in crowded and heterogeneous biological and synthetic soft matter environments, where controlled mobility of macromolecules is central to processes such as drug delivery, viral motion in mucus, and nanoparticle transport in gels.


\section*{Author Contributions}

S.M., J.B., P.S., and J.R.P. designed the research. S.M. carried out all simulations and analyzed the data. S.M., J.B., P.S., and J.R.P. wrote the manuscript. 

\section*{Acknowledgments}
We thank the funding and support from the IITB-Monash Research Academy and the Department of Biotechnology (DBT), India. We gratefully acknowledge the computational resources provided at the NCI National Facility systems at the Australian National University through the National Computational Merit Allocation Scheme supported by the Australian Government, the DUG-HPaaS, the MonARCH, and MASSIVE facilities maintained by Monash University.


\bibliographystyle{unsrt}
\bibliography{main}

\end{document}